\documentclass[twocolumn]{revtex4}
\usepackage{bm,amssymb,epsf}
\usepackage{color,colordvi}
\newcommand{\expec}[1]{{\rm E}\big(#1\big)}

\newcommand{\Dbra}[1]{\left\langle#1\right|}
\newcommand{\Dket}[1]{\left|#1\right\rangle}
\newcommand{\Qcommu}[2]{[#1,#2]}
\newcommand{\Qantico}[2]{\{#1,#2\}}

\newcommand{\gf}{|\hspace{.06em}0^{(0)}\rangle}
\newcommand{\gi}{|\hspace{.06em}0\rangle}
\newcommand{\gfa}{\langle 0^{(0)}\hspace{.06em}|}
\newcommand{\gia}{\langle 0\hspace{.06em}|}

\begin{document}
\bibliographystyle{apsrev}

\title{Kinetic Theory and Stochastic Simulation of Field Quanta}

\author{Hans Christian \"Ottinger}
\email[]{hco@mat.ethz.ch}
\homepage[]{http://www.polyphys.mat.ethz.ch/}
\affiliation{ETH Z\"urich, Department of Materials, Polymer Physics, HCI H 543,
CH-8093 Z\"urich, Switzerland}

\date{\today}

\begin{abstract}
We develop quantum electrodynamics into a kinetic-theory-like evolution equation for electrons, positrons and photons. To keep the ``collision rules'' simple, we make use of longitudinal and temporal photons in addition to the usual transverse photons. For our explicitly time-dependent approach, we introduce proper time-correlation functions. We then develop a stochastic simulation technique for solving the resulting kinetic equation and we comment on the subtleties associated with gauge invariance, Lorentz invariance and divergent integrals. To illustrate the validity and power of the proposed ideas, we show how a very simple simulation of the dynamics of field quanta can be used to obtain the main contribution to the anomalous magnetic moment of the electron.
\end{abstract}



\maketitle

\section{Introduction}
The goal of this work is to introduce particle-like simulations for quantum field theories. Rather than with particles in a classical sense, of course, we actually deal with field quanta. In the course of time, these field quanta undergo collisions of the types that we commonly represent by Feynman diagrams. We develop and illustrate the ideas in the context of quantum electrodynamics (QED), where the collisions are given by (i) emission or absorption of photons by electrons or positrons and (ii) electron-positron pair production or annihilation. The simulations to be developed consist of a stochastic sequence of such collisions.

Our particle-based simulations are supposed to be developed into an alternative to the widely used field-based simulations, which go back to K.\,G.~Wilson's famous formulation of lattice gauge theory \cite{Wilson74}. Computer simulations of lattice gauge theories with dynamic fermions \cite{DuaneKogut86,Gottliebetal87} have been established as a very successful tool in nonperturbative quantum field theory, but they are extremely demanding from a computational point of view (four-dimensional lattices, continuous evolution by hybrid-molecular-dynamics algorithms, conjugate-gradient calculations). It is hence interesting to explore complementary simulation ideas.

In Section~\ref{secfqf}, we introduce the ideas and notation of the four-photon approach to free electromagnetic fields and Dirac's spinor representation of free electrons and positrons. Interactions are then introduced in Section~\ref{secint}, where also the relativistic covariance and the gauge invariance of the theory are discussed. Proper correlation functions for our explicitly time-dependent approach are presented in Section~\ref{seccfmm}; in particular, we consider propagators, vertex functions and the electron form factor leading to the magnetic moment of the electron. We then have all the tools to develop a kinetic-theory-like simulation method for the field quanta in Section~\ref{secsim} and we present a particularly simple simulation that allows us to obtain the main contribution to the anomalous magnetic moment of the electron. A brief summary and a detailed discussion conclude the paper (Section~\ref{secsd}). Some useful definitions and relations are compiled in the Appendix.

Throughout this paper, we use natural units with $\hbar=c=\epsilon_0=1$ where $\hbar$ is the reduced Planck constant, $c$ is the speed of light, and $\epsilon_0$ is the electric permittivity. In these units, the electric charge is given by  $e_0 = \sqrt{4 \pi \alpha} \approx 0.30282212$, where $\alpha$ is the fine-structure constant. Only one further unit, naturally taken as mass or energy, remains to be specified.

\section{Free quantum fields}\label{secfqf}
We begin our development by introducing the free electromagnetic and electron-positron spinor fields. On the one hand, we lay out the basic notation and constructions, on the other hand, we offer a short but self-contained introduction to QED in a non-standard approach with longitudinal and temporal photons.

\subsection{Vector potential}\label{secfqfvp}
To avoid the need of considering Coulomb interactions explicitly, we use the four-photon quantization of the electromagnetic field. This elegant idea was originally developed in 1950 by Gupta \cite{Gupta50} for free electromagnetic fields and by Bleuler \cite{Bleuler50} in the presence of charged matter.

The four-vector potential with components $A_\mu$ is introduced as a Fourier transform in terms of polarization states and the corresponding creation and annihilation operators,
\begin{equation}\label{4vecpot}
    A(\bm{x}) = \frac{1}{\sqrt{(2\pi)^3}} \int \frac{d^3q}{\sqrt{2 q}}
    \left( n^\alpha_{\bm{q}} a^{\alpha \, \dag}_{\bm{q}}
    + \varepsilon_{\alpha} n^\alpha_{-\bm{q}} a^\alpha_{-\bm{q}} \right)
    e^{- i \bm{q} \cdot \bm{x}} .
\end{equation}
We denote the corresponding Fourier component in the integrand by
\begin{equation}\label{4vecpotFou}
    A_{\bm{q}} = \frac{1}{\sqrt{2 q}}
    \left( n^\alpha_{\bm{q}} a^{\alpha \, \dag}_{\bm{q}}
    + \varepsilon_{\alpha} n^\alpha_{-\bm{q}} a^\alpha_{-\bm{q}} \right) .
\end{equation}
The temporal unit vector,
\begin{equation}\label{polarization0}
    n^0_{\bm{q}} =
    \left( \begin{array}{c}
      1 \\
      0 \\
      0 \\
      0 \\
    \end{array} \right) ,
\end{equation}
is actually independent of $\bm{q}$. The three orthonormal spatial polarization vectors are chosen as
\begin{equation}\label{polarization1}
    n^1_{\bm{q}} = \frac{1}{\sqrt{q_1^2+q_2^2}}
    \left( \begin{array}{c}
      0 \\
      q_2 \\
      - q_1 \\
      0 \\
    \end{array} \right) ,
\end{equation}
\begin{equation}\label{polarization2}
    n^2_{\bm{q}} = \frac{1}{q \sqrt{q_1^2+q_2^2}}
    \left( \begin{array}{c}
      0 \\
      q_1 q_3 \\
      q_2 q_3\\
      - q_1^2 - q_2^2 \\
    \end{array} \right) ,
\end{equation}
and
\begin{equation}\label{polarization3}
    n^3_{\bm{q}} = \frac{1}{q}
    \left( \begin{array}{c}
      0 \\
      q_1\\
      q_2\\
      q_3\\
    \end{array} \right) .
\end{equation}
The polarization vectors $n^1_{\bm{q}}$ and $n^2_{\bm{q}}$ correspond to transverse photons, $n^3_{\bm{q}}$ corresponds to longitudinal photons. Note the symmetry property
\begin{equation}\label{polarsym}
    n^\alpha_{-\bm{q}} = (-1)^\alpha \, n^\alpha_{\bm{q}} .
\end{equation}
The sign $\varepsilon_{\alpha}$ (defined as $+1$ for transverse and longitudinal photons, $-1$ for temporal photons) leads to a non-self-adjoint nature of the vector potential (\ref{4vecpot}). The spatial components are self-adjoint, whereas the temporal component $A_0$ is anti-self-adjoint. The latter statement can be expressed as
\begin{equation}\label{antiHermitalt}
    A_0^\dag(\bm{x}) = - A_0(\bm{x}) = A^0(\bm{x}) ,
\end{equation}
where we have assumed a Minkowski metric $\eta^{\mu\nu}=\eta_{\mu\nu}$ with signature $(-,+,+,+)$, that is, $\eta^{00}=\eta_{00}=-1$. It is hence important to note that Eqs.~(\ref{4vecpot}) and (\ref{polarization0}) specify $A_0$ with a lower index. With these sign conventions, we further obtain a modified completeness relation for the polarization vectors,
\begin{equation}\label{modcomplete}
    \sum_\alpha \varepsilon_{\alpha} \, (n^\alpha_{\bm{q}})_\mu (n^\alpha_{\bm{q}})_\nu = \eta_{\mu\nu} .
\end{equation}

The four photon creation and annihilation operators, $a^{\alpha \, \dag}_{\bm{q}}$ and $a^\alpha_{\bm{q}}$, introduced in Eq.~(\ref{4vecpot}) satisfy the usual commutation relations,
\begin{equation}\label{acommutator}
    \Qcommu{a^\alpha_{\bm{q}}}{a^{\alpha' \, \dag}_{\bm{q}'}} = \delta_{\alpha\alpha'} \, \delta(\bm{q}-\bm{q}') .
\end{equation}
We can hence build the four-photon Fock space in the usual way by multiple application of all the operators $a^{\alpha \, \dag}_{\bm{q}}$ for all $\bm{q} \in \mathbb{R}^d$ on a ground state (which is annihilated by any $a^\alpha_{\bm{q}}$). The full Hilbert space factorizes into spaces obtained by repeated application of $a^{\alpha \, \dag}_{\bm{q}}$ for each mode $\bm{q}$ (see, for example, Secs.~1 and 2 of \cite{FetterWalecka} or Secs.~12.1 and 12.2 of \cite{BjorkenDrell} for more details on the construction of such Fock spaces).

There is only one modification of the standard construction: a modification of the scalar product, which we present as the construction of bra- from ket-vectors. If a basis vector of the ket space contains $n$ temporal photons, the corresponding standard bra-vector is multiplied by a factor $(-1)^n$. This minus sign associated with each temporal photon in the bra-space is introduced such that the vector potential (\ref{4vecpot}) becomes self-adjoint for the modified scalar product. The use of an indefinite scalar product may be alarming because it might endanger the probabilistic interpretation of the results as, for example, the norm of an odd-number temporal photon state is negative. However, no interpretation problems arise for physically admissible states, where admissibility is defined in terms of a proper version of the covariant Lorentz gauge condition. In particular, the Lorentz condition implies that physical states involve equal numbers of longitudinal and temporal photons having the same momentum (see Chapter 17 of \cite{Boyarkin1}), which gives us an idea of how interpretational problems are avoided. The physical states have been listed in an elementary way in (2.16) of the pioneering work \cite{Bleuler50}. An elegant way of characterizing the physical states has been given in Section V.C.3 of \cite{Cohenetcpp} by transforming from longitudinal and temporal to left (gauge) and right (droite) photons,
\begin{equation}\label{gphotondef}
    a^{\rm g}_{\bm{q}} = \frac{1}{\sqrt{2}} ( a^3_{\bm{q}} + a^0_{\bm{q}} ) ,
\end{equation}
and
\begin{equation}\label{dphotondef}
    a^{\rm d}_{\bm{q}} = \frac{i}{\sqrt{2}} ( a^3_{\bm{q}} - a^0_{\bm{q}} ) .
\end{equation}
Physical states can contain an arbitrary number of g-photons, but no d-photons.

Finally, the free Hamiltonian for our massless photons is given by
\begin{equation}\label{H0gendef}
    H_{\rm EM}^{(0)} = \int q \, a^{\alpha \, \dag}_{\bm{q}} a^\alpha_{\bm{q}} \, d^3q ,
\end{equation}
where all four photons are clearly treated on an equal footing. Their energy is given by the relativistic expression for massless particles. The free Hamiltonian (\ref{H0gendef}) is self-adjoint both for the standard and for the indefinite scalar product.

The observable magnetic field, as obtained from the rotation of the spatial part of the vector potential (\ref{4vecpot}), is given by the spatial components of
\begin{eqnarray}
    B(\bm{x}) &=& \frac{i}{\sqrt{2(2\pi)^3}} \int d^3q \, \sqrt{q} \,
    \big( n^1_{\bm{q}} a^{2 \, \dag}_{\bm{q}} - n^1_{-\bm{q}} a^2_{-\bm{q}}
    \qquad \nonumber\\
    && \hspace{4em} - n^2_{\bm{q}} a^{1 \, \dag}_{\bm{q}} + n^2_{-\bm{q}} a^1_{-\bm{q}} \big)
    e^{- i \bm{q} \cdot \bm{x}} .
\label{magnfield}
\end{eqnarray}
Note that the magnetic field involves only transverse photon operators. The observable Fourier components are associated with $a^{1 \, \dag}_{\bm{q}} - a^1_{-\bm{q}}$ and $a^{2 \, \dag}_{\bm{q}} + a^2_{-\bm{q}}$ which, in view of the symmetries (\ref{polarsym}), coincide with the transverse Fourier components of the vector potential in Eq.~(\ref{4vecpotFou}).

\subsection{Dirac fields and spinor properties}
According to Eq.~(13.59) of \cite{BjorkenDrell}, the Hamiltonian associated with the Dirac equation for the free electron/positron can be written as
\begin{equation}\label{H0Diracdef}
    H^{(0)}_{\rm e/p} = \int E_p \left( b^{\sigma \, \dag}_{\bm{p}} b^\sigma_{\bm{p}}
    + d^{\sigma \, \dag}_{\bm{p}} d^\sigma_{\bm{p}} \right) d^3p ,
\end{equation}
where $E_p = \sqrt{\bm{p}^2 + m^2}$ gives the energy of a relativistic particle with mass $m$ and momentum $\bm{p}$, and a summation over the possible spin values $\sigma =\pm 1/2$ is implied by the same index occurring twice. The operators $b^{\sigma \, \dag}_{\bm{p}}$ and $b^\sigma_{\bm{p}}$ create and annihilate an electron of momentum $\bm{p}$ and spin $\sigma$. Similarly, the operators $d^{\sigma \, \dag}_{\bm{p}}$ and $d^\sigma_{\bm{p}}$ create and annihilate a positron of momentum $\bm{p}$ and spin $\sigma$. As we are now interested in fermions instead of the bosonic photons, we need to specify the fundamental anticommutation relations for the creation and annihilation operators. All particle and antiparticle creation operators anticommute among each other, and so do the annihilation operators. The only nontrivial anticommutation relations are
\begin{equation}\label{banticommutator}
    \Qantico{b^\sigma_{\bm{p}}}{b^{\sigma' \, \dag}_{\bm{p}'}} =
    \delta_{\sigma\sigma'} \, \delta(\bm{p}-\bm{p}') ,
\end{equation}
and
\begin{equation}\label{danticommutator}
    \Qantico{d^\sigma_{\bm{p}}}{d^{\sigma' \, \dag}_{\bm{p}'}} =
    \delta_{\sigma\sigma'} \, \delta(\bm{p}-\bm{p}') .
\end{equation}

The four-component spinor field associated with annihilating an electron or creating a positron, each coming with the two values $\pm 1/2$ of the spin, has the Fourier representation (see Eq.~(13.50) of \cite{BjorkenDrell})
\begin{equation}\label{spinorfield}
    \psi(\bm{x}) = \int \frac{d^3p}{\sqrt{(2\pi)^3}} \, \sqrt{\frac{m}{E_p}}
    \left( v^{-\sigma}_{-\bm{p}}  d^{-\sigma \, \dag}_{-\bm{p}}
    + u^{\sigma}_{\bm{p}} b^{\sigma}_{\bm{p}} \right) e^{i \bm{p} \cdot \bm{x}} ,
\end{equation}
with the adjoint field operator creating an electron or annihilating a positron,
\begin{equation}\label{spinorfielddag}
    \psi^\dag(\bm{x}) = \int \frac{d^3p}{\sqrt{(2\pi)^3}} \, \sqrt{\frac{m}{E_p}}
    \left( v^{-\sigma *}_{-\bm{p}}  d^{-\sigma}_{-\bm{p}}
    + u^{\sigma *}_{\bm{p}} b^{\sigma \, \dag}_{\bm{p}} \right) e^{-i \bm{p} \cdot \bm{x}} ,
\end{equation}
where an asterisk implies both complex conjugation and transposition of a vector or matrix. Equation (\ref{spinorfielddag}) should be compared to Eq.~(\ref{4vecpot}); instead of the polarization vectors for photons, the spinors $u$ and $v$ occur for relativistic electrons and positrons. In the standard representation, they are given by
\begin{equation}\label{spinorsu1}
    u^{1/2}_{\bm{p}} = \sqrt{\frac{E_p+m}{2m}}
    \left( \begin{array}{c}
      1 \\
      0 \\
      \hat{p}_3 \\
      \hat{p}_1 + i \hat{p}_2 \\
    \end{array} \right) ,
\end{equation}
\begin{equation}\label{spinorsu2}
    u^{-1/2}_{\bm{p}} = \sqrt{\frac{E_p+m}{2m}}
    \left( \begin{array}{c}
      0 \\
      1 \\
      \hat{p}_1 - i \hat{p}_2 \\
      - \hat{p}_3 \\
     \end{array} \right) ,
\end{equation}
\begin{equation}\label{spinorsv1}
    v^{1/2}_{\bm{p}} = \sqrt{\frac{E_p+m}{2m}}
    \left( \begin{array}{c}
      \hat{p}_1 - i \hat{p}_2 \\
      - \hat{p}_3 \\
      0 \\
      1 \\
     \end{array} \right) ,
\end{equation}
and
\begin{equation}\label{spinorsv2}
    v^{-1/2}_{\bm{p}} = \sqrt{\frac{E_p+m}{2m}}
    \left( \begin{array}{c}
      \hat{p}_3 \\
      \hat{p}_1 + i \hat{p}_2 \\
      1 \\
      0 \\
    \end{array} \right) ,
\end{equation}
where we have introduced the notation $\hat{p}_j = p_j/(E_p+m)$ for conveniently normalized momenta $p_j=p^j$.

In the context of a relativistic theory, it is useful to introduce $\bar{\psi} = \psi^\dag \gamma^0$, where $\gamma^0$ is a diagonal $4\times 4$ matrix with the diagonal elements $(1,1,-1,-1)$, because $\bar{\psi}$ has a more natural Lorentz transformation behavior than $\psi^\dag$ (see Section 2.2 of \cite{BjorkenDrellQM}). In particular, $\bar{\psi}(\bm{x}) \psi(\bm{x})$ is a Lorentz scalar, whereas $\psi^\dag(\bm{x}) \psi(\bm{x})$ is not. With the corresponding definitions $\bar{u} = u^* \gamma^0$ and $\bar{v} = v^* \gamma^0$, Eq.~(\ref{spinorfielddag}) can be multiplied by $\gamma^0$ and rewritten as
\begin{equation}\label{spinorfieldbar}
    \bar{\psi}(\bm{x}) = \int \frac{d^3p}{\sqrt{(2\pi)^3}} \, \sqrt{\frac{m}{E_p}}
    \left( \bar{v}^{-\sigma}_{-\bm{p}}  d^{-\sigma}_{-\bm{p}}
    + \bar{u}^{\sigma}_{\bm{p}} b^{\sigma \, \dag}_{\bm{p}} \right) e^{-i \bm{p} \cdot \bm{x}} .
\end{equation}
We introduce the Fourier components in Eqs.~(\ref{spinorfield}) and (\ref{spinorfieldbar}) as
\begin{equation}\label{spinorfieldFou}
    \psi_{\bm{p}} = \sqrt{\frac{m}{E_p}}
    \left( v^{\sigma}_{\bm{p}}  d^{\sigma \, \dag}_{\bm{p}}
    + u^{-\sigma}_{-\bm{p}} b^{-\sigma}_{-\bm{p}} \right) ,
\end{equation}
and
\begin{equation}\label{spinorfieldbarFou}
    \bar{\psi}_{\bm{p}} = \sqrt{\frac{m}{E_p}}
    \left( \bar{v}^{-\sigma}_{-\bm{p}}  d^{-\sigma}_{-\bm{p}}
    + \bar{u}^{\sigma}_{\bm{p}} b^{\sigma \, \dag}_{\bm{p}} \right) .
\end{equation}

For $\bm{p}=\bm{0}$, the column vectors (\ref{spinorsu1})--(\ref{spinorsv2}) clearly form an orthogonal basis of the four-dimensional vector space characterizing the spins of electrons and positrons. For arbitrary $\bm{p}$, we still have orthogonality and completeness relations. Actually, there are two versions, one for spinors with equal $\bm{p}$ and one for spinors with opposite $\bm{p}$. For example, one form of the orthogonality relations reads
\begin{equation}\label{upsorthog}
    \bar{u}^{\sigma}_{\bm{p}} \, u^{\sigma'}_{\bm{p}} =
    -\bar{v}^{\sigma}_{\bm{p}} \, v^{\sigma'}_{\bm{p}} =
    \delta_{\sigma\sigma'} ,
\end{equation}
and
\begin{equation}\label{uvpsorthog}
    \bar{u}^{\sigma}_{\bm{p}} \, v^{\sigma'}_{\bm{p}} =
    \bar{v}^{\sigma}_{\bm{p}} \, u^{\sigma'}_{\bm{p}} = 0 .
\end{equation}
To establish the corresponding completeness relation, we consider the $4 \times 4$ matrices obtained as sums over tensor products of spinors and write them in a compact form in terms of the Pauli matrices (see appendix) and $2 \times 2$ unit matrices. We find
\begin{eqnarray}
    \Lambda_{\rm e}(\bm{p}) &=& \sum_\sigma u^{\sigma}_{\bm{p}} \bar{u}^{\sigma}_{\bm{p}} \nonumber \\
    &=& \frac{1}{2m}
    \left( \begin{array}{cc}
      (E_p+m)\bm{1} & - p_j \sigma^j \\
      p_j \sigma^j & - (E_p-m)\bm{1} \\
    \end{array} \right) \nonumber \\
    &=& \frac{1}{2m} \, (E_p \gamma^0 - p_j \gamma^j + m \, \bm{1}) \,,
\label{projectorele}
\end{eqnarray}
and
\begin{eqnarray}
    \Lambda_{\rm p}(\bm{p}) &=& - \sum_\sigma v^{\sigma}_{\bm{p}} \bar{v}^{\sigma}_{\bm{p}} \nonumber \\
    &=& \frac{1}{2m}
    \left( \begin{array}{cc}
      - (E_p-m)\bm{1} & p_j \sigma^j \\
      -p_j \sigma^j & (E_p+m)\bm{1} \\
    \end{array} \right) \nonumber \\
    &=& \frac{1}{2m} \, (- E_p \gamma^0 + p_j \gamma^j + m \, \bm{1}) \,,
\label{projectorpos}
\end{eqnarray}
where the Dirac $4 \times 4$ matrices $\gamma^j$ required for the compact reformulation in the last step are listed in the appendix. For $\bm{p}=\bm{0}$, we realize that $\Lambda_{\rm e}(\bm{p})$ and $\Lambda_{\rm p}(\bm{p})$ may be regarded as projectors to the electron and positron degrees of freedom, respectively. In Eqs.~(\ref{upsorthog})--(\ref{projectorpos}), there is no integration over $\bm{p}$ and we indicate the summations over $\sigma$ explicitly. Equations (\ref{projectorele}) and (\ref{projectorpos}) imply the completeness relation
\begin{equation}\label{projectorsum}
    \Lambda_{\rm e}(\bm{p}) + \Lambda_{\rm p}(\bm{p}) = \bm{1} ,
\end{equation}
and the additional property
\begin{equation}\label{projectordif}
    \Lambda_{\rm e}(\bm{p}) - \Lambda_{\rm p}(-\bm{p}) = \frac{E_p}{m} \, \gamma^0 .
\end{equation}

\section{Interactions}\label{secint}
The interaction between charged leptons and photons is given by the Hamiltonian
\begin{equation}\label{Hintexpression}
    H^{(1)} = - \int d^3x \, J^\mu(\bm{x}) A_\mu(\bm{x}) \, ,
\end{equation}
where we have made use of the electric charge density,
\begin{equation}\label{charge}
    J^0(\bm{x}) = -e_0 \psi^\dag(\bm{x}) \psi(\bm{x})
    = -e_0 \bar{\psi}(\bm{x}) \gamma^0 \psi(\bm{x}) \, ,
\end{equation}
and the electric current density,
\begin{equation}\label{current}
    J^j(\bm{x}) = -e_0 \bar{\psi}(\bm{x}) \gamma^j \psi(\bm{x}) \, .
\end{equation}
Equations~(\ref{charge}) and (\ref{current}) define a four-vector. Note that, in the standard scalar product, the Hamiltonian $H^{(1)}$ is not self-adjoint because $A_0(\bm{x})$ is anti-self-adjoint. However, in the indefinite scalar product, the Hamiltonian $H^{(1)}$ becomes self-adjoint. As suggested in Eq.~(13.61) of \cite{BjorkenDrell}, we actually use the normal-ordered versions of the charge and current densities (all creation operators are to the left of all annihilation operators in a product) to avoid irrelevant infinite contributions.

\subsection{Collision amplitudes}
The interaction term (\ref{Hintexpression}) involves two spinor fields and one vector potential, and hence such a ternary interaction corresponds to one photon and two lepton lines in a Feynman diagram. After inserting the Fourier transforms (\ref{4vecpot}), (\ref{spinorfield}) and (\ref{spinorfieldbar}), we write the interaction between Dirac and electromagnetic fields in terms of two contributions, $H^{(1)} = H^{(1)}_{\rm ea} + H^{(1)}_{\rm pp}$. The first contribution describes the spin dependent \emph{emission and absorption} of photons by electrons and positrons, the second term accounts for \emph{pair production} and annihilation. These contributions are of the following form,
\begin{eqnarray}
    H^{(1)}_{\rm ea} &=& \int d^3p \, d^3p'
    \left[ h^{\sigma \sigma' \mu}_{\bm{p} \bm{p}'} \, b^{\sigma\dag}_{\bm{p}} \, b^{\sigma'}_{\bm{p}'}
    - h^{\sigma' \sigma \mu}_{\bm{p}' \bm{p}}
     \, d^{-\sigma\dag}_{\bm{p}} \, d^{-\sigma'}_{\bm{p}'} \right]
    \nonumber\\ &\times&
    \big( n^\alpha_{\bm{p}'-\bm{p}} a^{\alpha\,\dag}_{\bm{p}'-\bm{p}}
    + \varepsilon_{\alpha} n^\alpha_{\bm{p}-\bm{p}'} a^{\alpha}_{\bm{p}-\bm{p}'} \big)_\mu \,,
\label{Hinteractea}
\end{eqnarray}
and
\begin{eqnarray}
    H^{(1)}_{\rm pp} &=& \int d^3p \, d^3p'
    \left[ \tilde{h}^{\sigma \sigma' \mu}_{\bm{p} \bm{p}'} \, d^{-\sigma}_{-\bm{p}} \, b^{\sigma'}_{\bm{p}'}
    + \tilde{h}^{\sigma' \sigma \mu}_{\bm{p}' \bm{p}}
     \, b^{\sigma'\dag}_{-\bm{p}'} \, d^{-\sigma\dag}_{\bm{p}} \right]
    \nonumber\\ &\times&
    \big( n^\alpha_{\bm{p}'-\bm{p}} a^{\alpha\,\dag}_{\bm{p}'-\bm{p}}
    + \varepsilon_{\alpha} n^\alpha_{\bm{p}-\bm{p}'} a^{\alpha}_{\bm{p}-\bm{p}'} \big)_\mu \,.
\label{Hinteractpp}
\end{eqnarray}
The complex collision amplitudes $h^{\sigma \sigma' \mu}_{\bm{p} \bm{p}'}$ and $\tilde{h}^{\sigma \sigma' \mu}_{\bm{p} \bm{p}'}$ are proportional to the electric charge $e_0$ and fully characterize electromagnetic interactions of photons with electrons and positrons. For the functional form of the collision amplitude $h^{\sigma \sigma' \mu}_{\bm{p} \bm{p}'}$ characterizing photon emission, we find
\begin{equation}\label{hphotemiss}
    h^{\sigma \sigma' \mu}_{\bm{p} \bm{p}'} =
    \frac{2 m \, e_0 \, \bar{u}^{\sigma}_{\bm{p}} \,
    \gamma^\mu \, u^{\sigma'}_{\bm{p}'}}{\sqrt{(4\pi)^3 E_p E_{p'} |\bm{p}-\bm{p}'|}} \, .
\end{equation}
The collision amplitude $\tilde{h}^{\sigma \sigma' \mu}_{\bm{p} \bm{p}'}$ for pair production is given by
\begin{equation}\label{hpairprod}
    \tilde{h}^{\sigma \sigma' \mu}_{\bm{p} \bm{p}'} =
    \frac{2 m \, e_0 \, \bar{v}^{-\sigma}_{-\bm{p}} \,
    \gamma^\mu \, u^{\sigma'}_{\bm{p}'}}{\sqrt{(4\pi)^3 E_p E_{p'} |\bm{p}-\bm{p}'|}} \, .
\end{equation}
These collision amplitudes have the following symmetries,
\begin{equation}\label{hphotemisssym}
    \left( h^{\sigma \sigma' \mu}_{\bm{p} \bm{p}'} \right)^* =
    h^{\sigma' \sigma \mu}_{\bm{p}' \bm{p}} ,
\end{equation}
and
\begin{equation}\label{hpairprodsym}
    \left( \tilde{h}^{\sigma \sigma' \mu}_{\bm{p} \bm{p}'} \right)^* =
    \tilde{h}^{\sigma' \sigma \mu}_{-\bm{p}' \, -\bm{p}} \,\, ,
\end{equation}
which are useful for verifying the adjointness properties of $H^{(1)}_{\rm ea}$ and $H^{(1)}_{\rm pp}$.

For the reader's convenience, the required spinor expressions are listed explicitly in Tables~\ref{tablecouplmatrix} and \ref{tablecouplmatriy} in the appendix. Also some symmetry relations are compiled in the appendix. Actually, the property (\ref{hphotemisssym}) follows from Eq.~(\ref{uuident}), and (\ref{hpairprodsym}) follows from Eqs.~(\ref{uvvuident}) and (\ref{uvident}). In deriving Eqs.~(\ref{Hinteractea}) and (\ref{Hinteractpp}), the symmetry properties (\ref{uuvvident}) and (\ref{uvvuident}) have been used, respectively.

\begin{widetext}

Let us consider the following (non-normalized) states consisting of $n$ electrons, $\bar{n}$ positrons and $\tilde{n}$ photons,
\begin{equation}\label{simstate}
    \left| \begin{array}{ccc}
      \sigma_1 \cdots \sigma_n & \bar{\sigma}_1 \cdots \bar{\sigma}_{\bar{n}}
               & \alpha_1 \cdots \alpha_{\tilde{n}} \\
      \bm{p}_1 \cdots \bm{p}_n & \bar{\bm{p}}_1 \cdots \bar{\bm{p}}_{\bar{n}}
               & \bm{q}_1 \cdots \bm{q}_{\tilde{n}} \\
    \end{array} \right\rangle
    = b^{\sigma_1 \, \dag}_{\bm{p}_1} \cdots b^{\sigma_n \, \dag}_{\bm{p}_n}
    d^{\bar{\sigma}_1 \, \dag}_{\bar{\bm{p}}_1} \cdots d^{\bar{\sigma}_{\bar{n}} \, \dag}_{\bar{\bm{p}}_{\bar{n}}}
    a^{\alpha_1 \, \dag}_{\bm{q}_1} \cdots a^{\alpha_{\tilde{n}} \, \dag}_{\bm{q}_{\tilde{n}}} \gf \,,
\end{equation}
where $\sum_{l=1}^{n} \bm{p}_l + \sum_{l=1}^{\bar{n}} \bar{\bm{p}}_l + \sum_{l=1}^{\tilde{n}} \bm{q}_l$ is the total momentum of the multi-particle state. These states are the base vectors of the Fock space that includes electrons and positrons in addition to photons and hence properly generalizes the Fock space of Section \ref{secfqfvp}. The states (\ref{simstate}) are eigenstates of the energy of the noninteracting system with eigenvalues $\sum_{l=1}^{n} E_{p_l} + \sum_{l=1}^{\bar{n}} E_{\bar{p}_l} + \sum_{l=1}^{\tilde{n}} q_l$. We now act with $H^{(1)} = H^{(1)}_{\rm ea}+H^{(1)}_{\rm pp}$ on a state (\ref{simstate}). Although the calculation is straightforward and the outcome is quite lengthy, we give the explicit result because it clarifies why we refer to a ``kinetic-theory-like'' approach and to ``collision amplitudes:''
\begin{eqnarray}
    H^{(1)} && \hspace{-1.6em} \left| \begin{array}{ccc}
      \sigma_1 \cdots \sigma_n & \bar{\sigma}_1 \cdots \bar{\sigma}_{\bar{n}}
               & \alpha_1 \cdots \alpha_{\tilde{n}} \\
      \bm{p}_1 \cdots \bm{p}_n & \bar{\bm{p}}_1 \cdots \bar{\bm{p}}_{\bar{n}}
               & \bm{q}_1 \cdots \bm{q}_{\tilde{n}} \\
    \end{array} \right\rangle
    = \nonumber \\
    && \sum_{j=1}^n \sum_{l=1}^{\tilde{n}}
    h^{\sigma \sigma_j \mu}_{\bm{p}_j+\bm{q}_l \, \bm{p}_j} \,
    \varepsilon_{\alpha_l} \left( n^{\alpha_l}_{\bm{q}_l} \right)_\mu
    \left| \begin{array}{ccc}
      \sigma_1 \cdots \sigma_{j-1} \hspace{1.4em} \sigma \hspace{1.4em} \sigma_{j+1} \cdots \sigma_n
               & \bar{\sigma}_1 \cdots \bar{\sigma}_{\bar{n}}
               & \alpha_1 \cdots \alpha_{l-1} \alpha_{l+1} \cdots \alpha_{\tilde{n}} \\
      \bm{p}_1 \cdots \bm{p}_{j-1} \, \bm{p}_j+\bm{q}_l \, \bm{p}_{j+1} \cdots \bm{p}_n
               & \bar{\bm{p}}_1 \cdots \bar{\bm{p}}_{\bar{n}}
               & \bm{q}_1 \cdots \bm{q}_{l-1} \, \bm{q}_{l+1} \cdots \bm{q}_{\tilde{n}} \\
    \end{array} \right\rangle
    \nonumber \\
    &-& \sum_{k=1}^{\bar{n}} \sum_{l=1}^{\tilde{n}}
    h^{-\bar{\sigma}_k \, -\sigma \, \mu}_{\bar{\bm{p}}_k \, \bar{\bm{p}}_k+\bm{q}_l} \,
    \varepsilon_{\alpha_l} \left( n^{\alpha_l}_{\bm{q}_l} \right)_\mu
    \left| \begin{array}{ccc}
      \sigma_1 \cdots \sigma_n & \bar{\sigma}_1 \cdots \bar{\sigma}_{k-1} \hspace{1.4em}
                 \sigma \hspace{1.4em} \bar{\sigma}_{k+1} \cdots \bar{\sigma}_{\bar{n}}
               & \alpha_1 \cdots \alpha_{l-1} \alpha_{l+1} \cdots \alpha_{\tilde{n}} \\
      \bm{p}_1 \cdots \bm{p}_n & \bar{\bm{p}}_1 \cdots \bar{\bm{p}}_{k-1} \,
                 \bar{\bm{p}}_k+\bm{q}_l \, \bar{\bm{p}}_{k+1} \cdots \bar{\bm{p}}_{\bar{n}}
               & \bm{q}_1 \cdots \bm{q}_{l-1} \, \bm{q}_{l+1} \cdots \bm{q}_{\tilde{n}} \\
    \end{array} \right\rangle
    \nonumber \\
    &-& \sum_{j=1}^n \sum_{k=1}^{\bar{n}} (-1)^{n+j+k} \,
    \tilde{h}^{-\bar{\sigma}_k \sigma_j \mu}_{-\bar{\bm{p}}_k \, \bm{p}_j} \,
    \left( n^\alpha_{\bm{p}_j+\bar{\bm{p}}_k} \right)_\mu
    \left| \begin{array}{ccc}
      \sigma_1 \cdots \sigma_{j-1} \, \sigma_{j+1} \cdots \sigma_n
               & \bar{\sigma}_1 \cdots \bar{\sigma}_{k-1} \, \bar{\sigma}_{k+1} \cdots \bar{\sigma}_{\bar{n}}
               & \alpha_1 \cdots \alpha_{\tilde{n}} \hspace{1.4em} \alpha \hspace{1.4em} \\
      \bm{p}_1 \cdots \bm{p}_{j-1} \, \bm{p}_{j+1} \cdots \bm{p}_n
               & \bar{\bm{p}}_1 \cdots \bar{\bm{p}}_{k-1} \, \bar{\bm{p}}_{k+1} \cdots \bar{\bm{p}}_{\bar{n}}
               & \bm{q}_1 \cdots \bm{q}_{\tilde{n}} \, \bm{p}_j+\bar{\bm{p}}_k \\
    \end{array} \right\rangle
    \nonumber \\
    &+& \sum_{j=1}^n \int d^3q \,
    h^{\sigma \sigma_j \mu}_{\bm{p}_j-\bm{q} \, \bm{p}_j} \, \left( n^\alpha_{\bm{q}} \right)_\mu
    \left| \begin{array}{ccc}
      \sigma_1 \cdots \sigma_{j-1} \hspace{1.4em} \sigma \hspace{1.4em} \sigma_{j+1} \cdots \sigma_n
               & \bar{\sigma}_1 \cdots \bar{\sigma}_{\bar{n}}
               & \alpha_1 \cdots \alpha_{\tilde{n}} \, \alpha \\
      \bm{p}_1 \cdots \bm{p}_{j-1} \, \bm{p}_j-\bm{q} \, \bm{p}_{j+1} \cdots \bm{p}_n
               & \bar{\bm{p}}_1 \cdots \bar{\bm{p}}_{\bar{n}}
               & \bm{q}_1 \cdots \bm{q}_{\tilde{n}} \, \, \bm{q} \\
    \end{array} \right\rangle
    \nonumber \\
    &-& \sum_{k=1}^{\bar{n}} \int d^3q \,
    h^{-\bar{\sigma}_k \, -\sigma \, \mu}_{\bar{\bm{p}}_k \, \bar{\bm{p}}_k-\bm{q}} \,
    \left( n^\alpha_{\bm{q}} \right)_\mu
    \left| \begin{array}{ccc}
      \sigma_1 \cdots \sigma_n & \bar{\sigma}_1 \cdots \bar{\sigma}_{k-1} \hspace{1.3em}
                 \sigma \hspace{1.3em} \bar{\sigma}_{k+1} \cdots \bar{\sigma}_{\bar{n}}
               & \alpha_1 \cdots \alpha_{\tilde{n}} \, \alpha \\
      \bm{p}_1 \cdots \bm{p}_n & \bar{\bm{p}}_1 \cdots \bar{\bm{p}}_{k-1} \,
                 \bar{\bm{p}}_k-\bm{q} \, \bar{\bm{p}}_{k+1} \cdots \bar{\bm{p}}_{\bar{n}}
               & \bm{q}_1 \cdots \bm{q}_{\tilde{n}} \, \, \bm{q} \\
    \end{array} \right\rangle
    \nonumber \\
    &+& \sum_{l=1}^{\tilde{n}} (-1)^{\bar{n}} \int d^3p \, \,
    \tilde{h}^{\sigma \, -\bar{\sigma} \, \mu}_{-\bm{p} \, \bm{q}_l-\bm{p}} \,
    \varepsilon_{\alpha_l} \left( n^{\alpha_l}_{\bm{q}_l} \right)_\mu
    \left| \begin{array}{ccc}
      \sigma_1 \cdots \sigma_n \, \, \sigma
               & \bar{\sigma}_1 \cdots \bar{\sigma}_{\bar{n}} \hspace{1.2em} \bar{\sigma} \hspace{1.0em}
               & \alpha_1 \cdots \alpha_{l-1} \alpha_{l+1} \cdots \alpha_{\tilde{n}} \\
      \bm{p}_1 \cdots \bm{p}_n \, \bm{p}
               & \bar{\bm{p}}_1 \cdots \bar{\bm{p}}_{\bar{n}} \, \bm{q}_l-\bm{p}
               & \bm{q}_1 \cdots \bm{q}_{l-1} \, \bm{q}_{l+1} \cdots \bm{q}_{\tilde{n}} \\
    \end{array} \right\rangle
    \nonumber \\
    &-& \sum_{j=1}^n \sum_{k=1}^{\bar{n}} \sum_{l=1}^{\tilde{n}}
    (-1)^{n+j+k} \, \delta(\bm{p}_j+\bar{\bm{p}}_k+\bm{q}_l) \,
    \tilde{h}^{-\bar{\sigma}_k \sigma_j \mu}_{-\bar{\bm{p}}_k \, \bm{p}_j} \,
    \varepsilon_{\alpha_l} \left( n^{\alpha_l}_{\bm{q}_l} \right)_\mu
    \nonumber \\ && \hspace{7em} \times \,
    \left| \begin{array}{ccc}
      \sigma_1 \cdots \sigma_{j-1} \, \sigma_{j+1} \cdots \sigma_n
               & \bar{\sigma}_1 \cdots \bar{\sigma}_{k-1} \, \bar{\sigma}_{k+1} \cdots \bar{\sigma}_{\bar{n}}
               & \alpha_1 \cdots \alpha_{l-1} \alpha_{l+1} \cdots \alpha_{\tilde{n}} \\
      \bm{p}_1 \cdots \bm{p}_{j-1} \, \bm{p}_{j+1} \cdots \bm{p}_n
               & \bar{\bm{p}}_1 \cdots \bar{\bm{p}}_{k-1} \, \bar{\bm{p}}_{k+1} \cdots \bar{\bm{p}}_{\bar{n}}
               & \bm{q}_1 \cdots \bm{q}_{l-1} \, \bm{q}_{l+1} \cdots \bm{q}_{\tilde{n}} \\
    \end{array} \right\rangle
    \nonumber \\
    \rule{0em}{6ex}
    &+& (-1)^{\bar{n}} \int d^3p \int d^3\bar{p} \, \,
    \tilde{h}^{\sigma \, -\bar{\sigma} \, \mu}_{-\bm{p} \, \bar{\bm{p}}} \,
    \left( n^\alpha_{-\bm{p}-\bar{\bm{p}}} \right)_\mu
    \left| \begin{array}{ccc}
      \sigma_1 \cdots \sigma_n \, \, \sigma
               & \bar{\sigma}_1 \cdots \bar{\sigma}_{\bar{n}} \, \, \bar{\sigma}
               & \alpha_1 \cdots \alpha_{\tilde{n}} \hspace{1.6em} \alpha \hspace{1.3em} \\
      \bm{p}_1 \cdots \bm{p}_n \, \bm{p}
               & \bar{\bm{p}}_1 \cdots \bar{\bm{p}}_{\bar{n}} \, \bar{\bm{p}}
               & \bm{q}_1 \cdots \bm{q}_{\tilde{n}} \, -\bm{p}-\bar{\bm{p}} \\
    \end{array} \right\rangle \,.
\label{Hsimstate}
\end{eqnarray}

\end{widetext}
Equation (\ref{Hsimstate}) summarizes the eight collision rules of QED for electrons, positrons and photons. The first three terms \emph{decrease} the number of particles \emph{by one}, either through absorption of a photon by an electron or positron or through annihilation of an electron-positron pair into a photon. These terms contain double sums corresponding to the selection of two particles (to be replaced by one) and do not involve integrations. The next three terms \emph{increase} the number of particles \emph{by one}, either through emission of a photon by an electron or positron or through creation of an electron-positron pair from a photon. The hallmark of these terms is a single sum corresponding to the selection of a disappearing particle (to be replaced by two) and a momentum integration for one of the two new particles. The last two terms \emph{decrease} or \emph{increase} the number of particles \emph{by three}, an electron, a positron and a photon in either case. They contain a triple sum, corresponding to selecting three particles, and a double integral, corresponding to the freedom of choosing two momenta, respectively. Momentum conservation is respected by all of the eight terms, and each of the terms is proportional to a collision amplitude $h$ or $\tilde{h}$. These ``collision rules'' are the core of the simulation of QED to be developed below.

\subsection{Lorentz and gauge invariance}
After defining the Hilbert space, the fields and the various contributions to the Hamiltonian, we should now ask whether the resulting theory has the physically required invariance properties. The proper relativistic form is not at all obvious because the Hamiltonian approach assigns a special role to time. In the Hamiltonians (\ref{H0gendef}) and (\ref{H0Diracdef}) for the free electromagnetic and electron/positron fields, the relativistic energy-momentum relations for massless and massive particles provide the proper ingredient. For the interaction Hamiltonian (\ref{Hintexpression}) to be Lorentz invariant, $A_\mu$ and $J^\mu$ have to be Lorentz four-vectors. For $A_\mu$, this is nicely achieved by the four-photon approach which puts temporal and spatial components on an equal footing, except for the sign of the Minkowski metric properly arising from the use of an indefinite scalar product. For $J^\mu$, Lorentz transformation behavior is achieved by construction of the spinors, which are chosen to be trivial for particles at rest and then need to have a suitable momentum dependence for moving particles. The correct relativistic transformation behavior is hence hidden in the collision amplitudes (\ref{hphotemiss}) and (\ref{hpairprod}).

By introducing four types of photons, we are dealing with more than the physically observable degrees of freedom. In other words, our Hamiltonian dynamics takes place in an unphysically large space. The benefits are the simplicity and elegance of the equations in the larger space. The drawback is the need to identify the physical observables. As we rely on the approach of Gupta and Bleuler, this problem has been solved by introducing left and right photons, where the latter correspond to unphysical degrees of freedom. We here play safe and consider only correlation functions involving the transverse photons directly related to the observable magnetic field.

One might be worried about the hidden character of fundamental physical invariance properties. However, experience with the standard form of the Maxwell equations for electromagnetic fields and the usefulness of conveniently fixed potentials (fixed gauge) clearly underlines that such a procedure is very common and absolutely unproblematic.

\section{Correlation functions and magnetic moment}\label{seccfmm}
As we do not treat time and space on an equal footing and as we work in a particular gauge, it is important to introduce the appropriate correlation functions carefully. We here introduce and discuss some useful time-ordered two- and three-time correlation functions.

\subsection{Two- and three-time correlations}
We first consider the two-time correlation function of two operators $A$ and $B$. Actually, it is more convenient to study Fourier transforms. When time-ordering is properly taken into account, the correlation in the frequency domain is given by
\begin{equation}\label{correlFT2}
    {\cal C}^{(2)}_\omega(A,B) = \Dbra{0} A R_\omega B \Dket{0} \pm
    \Dbra{0} B R_{-\omega} A \Dket{0} \,,
\end{equation}
where we have introduced the evolution operator
\begin{equation}\label{Romdef}
    R_\omega = i \int_0^\infty dt \, e^{-i H t + i \omega t - \epsilon t}
    = (H - \omega - i \epsilon)^{-1} .
\end{equation}
The plus or minus sign in the definition (\ref{correlFT2}) depends on how $A$ and $B$ are related to commuting or anticommuting operators, the small parameter $\epsilon$ ensures the convergence of the time integral. Note that $R_\omega$ involves the full Hamiltonian and possesses the perturbation expansion,
\begin{eqnarray}
    R_{\omega} &=& \big[ 1 + R^{(0)}_{\omega} H^{(1)} \big]^{-1} R^{(0)}_{\omega} \nonumber\\
    &=& R^{(0)}_{\omega} - R^{(0)}_{\omega} H^{(1)} R^{(0)}_{\omega}
    + R^{(0)}_{\omega} H^{(1)} R^{(0)}_{\omega} H^{(1)} R^{(0)}_{\omega} \nonumber\\
    && - \ldots ,
\label{Romperturb}
\end{eqnarray}
where $R^{(0)}_{\omega}$ is defined as $R_\omega$ in Eq.~(\ref{Romdef}), but with the free Hamiltonian $H^{(0)}$ instead of $H$. The simplicity of this perturbation expansion for $R_\omega$ actually provides the motivation for considering Fourier transforms. A full perturbation theory of the correlation functions (\ref{correlFT2}) requires an additional expansion of the vacuum states. It is given by the following relations between the vacuum states of the free and interacting theories,
\begin{eqnarray}
    \gi  &=& \big[ 1 + R^{(0)}_0 H^{(1)} \big]^{-1} \gf , \nonumber\\
    \gia &=& \gfa \big[ 1 + H^{(1)} R^{(0)}_0 \big]^{-1} .
\label{vacuumexpansion}
\end{eqnarray}
To obtain this result, it is important to realize that $H^{(1)}$ is self-adjoint for the modified, indefinite scalar product. In the limit of a vanishing regularization parameter, the operators $R^{(0)}_{\omega}$ are self-adjoint for both the modified and the standard scalar products. Note the similarity of the perturbation expansions of the vacuum state and of $R_{\omega}$ for $\omega=0$.

For time-ordered three-time correlation functions, the proper Fourier transform can be expressed as
\begin{eqnarray}
    {\cal C}^{(3)}_{\omega',\omega}(A,B,C) &=& \Dbra{0} A R_{\omega'} B R_\omega C \Dket{0}
    \nonumber \\
    &-& \Dbra{0} C R_{-\omega} B R_{-\omega'} A \Dket{0}
    \nonumber \\
    &-& \Dbra{0} C R_{-\omega} A R_{\omega'-\omega} B \Dket{0}
    \nonumber \\
    &+& \Dbra{0} A R_{\omega'} C R_{\omega'-\omega} B \Dket{0}
    \nonumber \\
    &+& \Dbra{0} B R_{\omega-\omega'} A R_{\omega} C \Dket{0}
    \nonumber \\
    &-& \Dbra{0} B R_{\omega-\omega'} C R_{-\omega'} A \Dket{0}
    \,, \qquad
\label{correlFT3}
\end{eqnarray}
where the signs have been chosen for the case that $A$ and $C$ are fermion operators, whereas $B$ is a boson operator. Note that the frequency $\omega'$ occurs all the way between $A$ and $B$, and $\omega$ between $B$ and $C$. If the order of $A$ and $B$ (or $B$ and $C$) is reversed, the frequency $\omega'$ (or $\omega$) picks up a minus sign.

\subsection{Propagators}
The standard form of Feynman's fermion propagator $S_{\alpha\beta}$ in Fourier space can be introduced by choosing $A$ and $B$ in Eq.~(\ref{correlFT2}) as Fourier components (\ref{spinorfieldFou}) and (\ref{spinorfieldbarFou}) of the fermion field. More precisely, we define
\begin{equation}\label{stanfermpropFou}
    {\cal C}^{(2)}_\omega(\psi_{\alpha,\bm{p}'},\bar{\psi}_{\beta,\bm{p}}) =
    - S_{\alpha\beta}(\omega,\bm{p}) \, \delta(\bm{p}+\bm{p}') ,
\end{equation}
where, for fermion operators, the minus sign must be used in the definition (\ref{correlFT2}). In space-time, this corresponds to the usual definition in terms of a time-ordered product (see, for example, Eq.~(13.72) of \cite{BjorkenDrell} or p.\,112 of \cite{Zee}),
\begin{eqnarray}
    i \tilde{S}_{\alpha\beta}(t,\bm{x}) &=& \Theta(t) \, \Dbra{0} \psi_\alpha(\bm{x}) e^{-i H t} \bar{\psi}_\beta(\bm{0}) \Dket{0} \nonumber \\
    &-& \Theta(-t) \, \Dbra{0} \bar{\psi}_\beta(\bm{0}) e^{i H t} \psi_\alpha(\bm{x}) \Dket{0} \,.
    \qquad
\label{stanfermprop}
\end{eqnarray}
In a more compact notation, Eq.~(\ref{stanfermprop}) can be expressed as
\begin{equation}\label{stanfermprop4}
    i \tilde{S}_{\alpha\beta}(x) = \Dbra{0} T[\psi_\alpha(x) \bar{\psi}_\beta(0)] \Dket{0} \,,
\end{equation}
where $x=(t,\bm{x})$ and $0=(0,\bm{0})$ are four-vectors and $T$ is the time-ordering operation.

For the free theory, it is straightforward to evaluate the correlation function in Eq.~(\ref{stanfermpropFou}) for the Fourier components given in Eqs.~(\ref{spinorfieldFou}) and (\ref{spinorfieldbarFou}). We obtain the free propagator
\begin{eqnarray}
    S^{(0)}(\omega,\bm{p}) &=& - \frac{m}{E_p} \left[
    \frac{\Lambda_{\rm e}(\bm{p})}{E_p - \omega - i \epsilon}
    + \frac{\Lambda_{\rm p}(-\bm{p})}{E_p + \omega - i \epsilon} \right] \nonumber \\
    &=& \frac{\omega \gamma^0 - p_j \gamma^j + m \bm{1}}{\omega^2 - \bm{p}^2 - m^2 + i \epsilon} \,.
\label{stanfermpropFou40}
\end{eqnarray}
In the second step, we have used Eqs.~(\ref{projectorele}) an (\ref{projectorpos}) to obtain the standard form of the free fermion propagator.

A photon propagator can be introduced  in the same way as the fermion propagator (\ref{stanfermpropFou}), but now with the plus sign for boson operators in the definition (\ref{correlFT2}),
\begin{equation}\label{stanphotpropFou}
    {\cal C}^{(2)}_\omega(A_{\mu,\bm{q}'},A_{\nu,\bm{q}}) =
    - D_{\mu\nu}(\omega,\bm{q}) \, \delta(\bm{q}+\bm{q}') ,
\end{equation}
From a straightforward calculation, we obtain the free photon propagator
\begin{equation}\label{stanphotpropFou40}
    D^{(0)}_{\mu\nu}(\omega,\bm{q}) =
    \frac{\eta_{\mu\nu}}{\omega^2 - q^2 + i \epsilon} \,.
\end{equation}
This result illustrates that the four-photon approach corresponds to a particularly convenient gauge.

\subsection{Vertex functions}
As important further correlation functions, we introduce the vertex functions $\Gamma^\nu$ through special cases of a three-time correlation function,
\begin{eqnarray}
    {\cal C}^{(3)}_{\omega' \omega}(\psi_{\alpha,-\bm{p}'}, A_{\mu,\bm{q}}, \bar{\psi}_{\beta,\bm{p}}) &=&
    \frac{e_0}{\sqrt{(2\pi)^3}} \, \delta(\bm{p}+\bm{q}-\bm{p}') \nonumber \\
    && \hspace{-11em} \times \, D_{\mu\nu}(\omega'-\omega,\bm{p}'-\bm{p}) \nonumber \\
    && \hspace{-11em} \times \, \Big[ S(\omega',\bm{p}') \,
    \Gamma^\nu(\omega',\bm{p}',\omega,\bm{p}) \, S(\omega,\bm{p}) \Big]_{\alpha\beta} \,, \qquad
\label{deffvertex}
\end{eqnarray}
where the $\Gamma^\nu$ are $4 \times 4$ matrices in spinor space, just like the fermion propagators $S$. In the compact four-vector notation, we can write
\begin{eqnarray}
    {\cal C}^{(3)}_{-\omega' \omega}(\psi_{\alpha,\bm{p}'}, A_{\mu,\bm{q}}, \bar{\psi}_{\beta,\bm{p}}) &=&
    - \frac{1}{(2\pi)^{11/2}} \int dq^0 \nonumber \\
    && \hspace{-8em} \times \, \int d^4x \, d^4x' \, d^4y \,
    e^{i(p \cdot x + p' \cdot x' + q \cdot y)} \nonumber \\
    && \hspace{-8em} \times \,
    \Dbra{0} T[\psi_\alpha(x') A_\mu(y) \bar{\psi}_\beta(x)] \Dket{0} \,,
\label{deffvertex4}
\end{eqnarray}
where we have made use of the following four-vectors: $x=(t,\bm{x})$, $x'=(t',\bm{x}')$, $y=(t'',\bm{y})$, $p=(\omega,\bm{p})$, $p'=(\omega',\bm{p}')$, and $q=(q^0,\bm{q})$ so that, for example, $p \cdot x = \bm{p} \cdot \bm{x} - \omega t$.

With the propagators $S(\omega,\bm{p})$, $D(\omega,\bm{q})$ and the vertex functions $\Gamma^\nu(\omega',\bm{p}',\omega,\bm{p})$ defined in Eqs.~(\ref{stanfermpropFou}), (\ref{stanphotpropFou}) and (\ref{deffvertex}), we have now introduced the most important correlation functions QED.

\subsection{Perturbation theory}
\begin{figure}
\centerline{\epsfxsize=4cm \epsffile{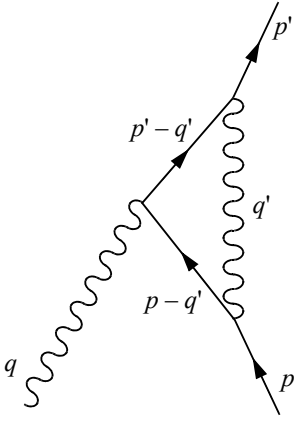}}
\caption[ ] {Feynman diagram corresponding to the contribution ${\Gamma^\mu}^{(3)}$ to the vertex function.}
\label{qed_Feynman}
\end{figure}

To gain a better understanding of the vertex functions $\Gamma^\nu(\omega',\bm{p}',\omega,\bm{p})$, we would like to calculate the lowest-order contributions by perturbation theory. As the definition (\ref{deffvertex}) involves three fields, the vertex functions vanish for the free theory. Only odd orders of perturbation theory can contribute. Even the first-order result,
\begin{equation}\label{vertexfct1}
    {\Gamma^\mu}^{(1)} = \gamma^\mu ,
\end{equation}
is already tedious to get within our approach, which is highly inappropriate for constructing perturbation expansions. It is hence recommended that perturbation theory should be based on compact expressions of the form given in Eqs.~(\ref{stanfermprop4}) and (\ref{deffvertex4}), for which elegant and powerful methods based on Feynman diagrams are available in any textbook on quantum field theory. Following the standard rules (see, for example, Appendix~B of \cite{BjorkenDrell} or Appendix~C of \cite{Zee} with the explicit result given in Eq.~(III.6.8) on p.\,197) we obtain
\begin{eqnarray}
    {\Gamma^\mu}^{(3)}(\omega',\bm{p}',\omega,\bm{p}) &=& \int \frac{d^4q'}{(2\pi)^4} \,
    i D^{(0)}_{\nu'\nu}(q') \nonumber \\
    && \hspace{-8em} \times \,
    i e_0 \gamma^{\nu'} \, i S^{(0)}(p'-q') \, \gamma^\mu \, i S^{(0)}(p-q') \, i e_0 \gamma^\nu ,
    \qquad
\label{vertexfct3}
\end{eqnarray}
where this contribution results from the Feynman diagram shown in Fig.~\ref{qed_Feynman}.

By inserting the first part of Eq.~(\ref{stanfermpropFou40}) as well as Eq.~(\ref{stanphotpropFou40}) and using Cauchy's integral formula to perform the frequency integration over ${q'}^0$, we can rewrite the expression (\ref{vertexfct3}) as a three-dimensional integral,
\begin{eqnarray}
    {\Gamma^\mu}^{(3)}(\omega',\bm{p}',\omega,\bm{p}) &=& \int \frac{d^3q'}{(2\pi)^3} \,
    \frac{e_0^2 \, m^2}{2 q' E_{|\bm{p}'-\bm{q}'|} E_{|\bm{p}-\bm{q}'|}} \, \eta_{\nu\nu'} \quad \nonumber \\
    && \hspace{-8em} \times \, \bigg[ \,
    \frac{\gamma^{\nu'} \, \Lambda_{\rm e}(\bm{p}'-\bm{q}') \, \gamma^\mu \, \Lambda_{\rm e}(\bm{p}-\bm{q}')
    \, \gamma^\nu}{(E_{|\bm{p}'-\bm{q}'|}+q'-\omega')(E_{|\bm{p}-\bm{q}'|}+q'-\omega)}
    \nonumber \\
    && \hspace{-8em} + \,
    \frac{\gamma^{\nu'} \, \Lambda_{\rm p}(\bm{q}'-\bm{p}') \, \gamma^\mu \, \Lambda_{\rm p}(\bm{q}'-\bm{p})
    \, \gamma^\nu}{(E_{|\bm{p}'-\bm{q}'|}+q'+\omega')(E_{|\bm{p}-\bm{q}'|}+q'+\omega)}
    \nonumber \\
    && \hspace{-8em} + \, \left( \frac{1}{E_{|\bm{p}'-\bm{q}'|}+q'-\omega'}
    + \frac{1}{E_{|\bm{p}-\bm{q}'|}+q'+\omega} \right)
    \nonumber \\
    && \hspace{-4em} \times \,
    \frac{\gamma^{\nu'} \, \Lambda_{\rm e}(\bm{p}'-\bm{q}') \, \gamma^\mu \, \Lambda_{\rm p}(\bm{q}'-\bm{p})
    \, \gamma^\nu}{E_{|\bm{p}'-\bm{q}'|}+E_{|\bm{p}-\bm{q}'|}-\omega'+\omega}
    \nonumber \\
    && \hspace{-8em} + \, \left( \frac{1}{E_{|\bm{p}'-\bm{q}'|}+q'+\omega'}
    + \frac{1}{E_{|\bm{p}-\bm{q}'|}+q'-\omega} \right)
    \nonumber \\
    && \hspace{-4em} \times \,
    \frac{\gamma^{\nu'} \, \Lambda_{\rm p}(\bm{q}'-\bm{p}') \, \gamma^\mu \, \Lambda_{\rm e}(\bm{p}-\bm{q}')
    \, \gamma^\nu}{E_{|\bm{p}'-\bm{q}'|}+E_{|\bm{p}-\bm{q}'|}+\omega'-\omega} \, \bigg] \,.
\label{vertexfct33}
\end{eqnarray}
One still recognizes the importance of fermion propagators, but now split into electron and positron contributions.

\subsection{Form factors and magnetic moment}
Perturbation theory suggests that $\Gamma^\nu$ can be regarded as an effective interaction vertex when photons and electron-positron pairs are around during the basic ternary interaction process described by $\gamma^\nu$. Whereas $\gamma^\nu$ characterizes the electromagnetic interaction of a structureless point electron, the full interaction including a cloud of photons and electron-positron pairs around the electron during interaction with an external photon is characterized by $\Gamma^\nu$. The physics of the effective interactions can be described elegantly in terms of the electromagnetic form factors $F$ and $G$, which occur in the most general representation of four-vectors (see, for example, Section~10.6 of \cite{WeinbergQFT1} or Section~III.6 of \cite{Zee}),
\begin{eqnarray}
    \bar{u}^{\sigma'}_{\bm{p}'} \, {\Gamma^\nu}(E_{p'},\bm{p}',E_p,\bm{p}) \, u^{\sigma}_{\bm{p}} &=&
    \nonumber \\
    && \hspace{-10em} F(x) \, \bar{u}^{\sigma'}_{\bm{p}'} \, \gamma^\nu \, u^{\sigma}_{\bm{p}}
    + G(x) \, \frac{(\bm{p} + \bm{p}')^\nu}{2m} \,
    \bar{u}^{\sigma'}_{\bm{p}'} \, u^{\sigma}_{\bm{p}} \,, \qquad
\label{formfacintro}
\end{eqnarray}
where the real functions $F(x)$ and $G(x)$ depend on
\begin{equation}\label{momtransf}
    x^2 = 2 \, ( E_p E_{p'} - \bm{p} \cdot \bm{p}' - m^2 ) .
\end{equation}
They satisfy the normalization condition $F(0)+G(0)=1$, which is consistent with the result of first-order perturbation theory, that is, with the constant form factors $F(x)=1$ and $G(x)=0$. In view of the definition (\ref{hphotemiss}), it is natural to define the effective collision amplitude
\begin{equation}\label{hphotemisseff}
    {\cal H}^{\sigma' \sigma \mu}_{\bm{p}' \bm{p}} =
    \frac{2 m \, e_0 \, \bar{u}^{\sigma'}_{\bm{p}'} \,
    {\Gamma^\mu}(E_{p'},\bm{p}',E_p,\bm{p}) \,
    u^{\sigma}_{\bm{p}}}{\sqrt{(4\pi)^3 E_p E_{p'} |\bm{p}-\bm{p}'|}} \, .
\end{equation}

The magnetic moment, which characterizes the interaction of a charged particle with an electromagnetic field, is given by the $g$-factor and can be expressed in terms of the form factor (see, for example, Section~10.6 of \cite{WeinbergQFT1} or Section~III.6 of \cite{Zee})
\begin{equation}\label{gfacelectron}
    g = 2 F(0) = 2 - 2 G(0) \, .
\end{equation}
In lowest-order perturbation theory, we find $g=2$, whereas the experimental result is slightly larger,
\begin{equation}\label{gfacelectronexp}
    g \approx 2.002319 \, ,
\end{equation}
where significantly more accurate experimental results would be available (the relative experimental error for $g$ is smaller than $10^{-12}$). The celebrated approximate one-loop result of Schwinger \cite{Schwinger48} corresponding to the Feynman diagram in Fig.~\ref{qed_Feynman} is
\begin{equation}\label{gfacelectron1}
    g = 2 + \frac{\alpha}{\pi} = 2 + \frac{e_0^2}{4 \pi^2} \approx 2.002323 \, .
\end{equation}
Our goal is to find the anomalous $g$-factor deviating from $2$ by kinetic-theory-like simulations.

\section{Simulations}\label{secsim}
After introducing the proper formulation of Hamiltonians and correlation functions in the four-photon approach, we are now in a position to develop a kinetic-theory-like simulation methodology. The eight collision rules in Eq.~(\ref{Hsimstate}), together with the collision amplitudes (\ref{hphotemiss}) and (\ref{hpairprod}), provide the corner stones for the stochastic simulation technique to be developed here.

\subsection{Basic ideas}
The simulations proposed in the present paper are strongly inspired by the powerful stochastic simulation techniques for quantum master equations \cite{GardinerZoller,BreuerPetru,hco202,hco205}. Instead of looking for a deterministic solution of the Schr\"odinger equation
\begin{equation}\label{Schroedeq}
    \frac{d\Dket{\psi_t}}{dt} = - i ( H^{(0)} + H^{(1)} ) \Dket{\psi_t} ,
\end{equation}
with $H^{(0)} = H_{\rm EM}^{(0)}+H_{\rm e/p}^{(0)}$ and $H^{(1)} = H^{(1)}_{\rm ea} + H^{(1)}_{\rm pp}$, we consider a stochastic process $\Dket{\psi_t}$ in Hilbert space such that the expectation $\expec{\Dket{\psi_t}}$ satisfies Eq.~(\ref{Schroedeq}). Let us consider an important example of a stochastic process $\Dket{\psi_t}$. In that example, the process starts from a deterministic initial condition $\Dket{\psi_0}$ and continuously evolves according to the Schr\"odinger equation (\ref{Schroedeq}) for the free Hamiltonian in the absence of interactions,
\begin{equation}\label{Schroedeqfree}
    \frac{d\Dket{\psi_t}}{dt} = - i H^{(0)} \Dket{\psi_t} .
\end{equation}
Instead of having a continuously acting Hamiltonian $H^{(1)}$, we introduce a rate parameter $r$ that determines the occurrence of random times at which the following jump occurs,
\begin{equation}\label{Schroedeqjump}
    \Dket{\psi_t} \rightarrow \Dket{\psi_t} - \frac{1}{r} \, i H^{(1)} \Dket{\psi_t} .
\end{equation}
The expectation of the stochastic process with continuous evolution (\ref{Schroedeqfree}) and random jumps (\ref{Schroedeqjump}) occurring with rate $r$ solves the Schr\"odinger equation (\ref{Schroedeq}). The inverse rate $1/r$ plays the role of a time step that, however, occurs randomly according to an exponentially decaying distribution with average $1/r$. For large $r$, we recover an almost continuous evolution with small jumps. For small $r$, we have rare but large jumps according to the prefactor $1/r$ in Eq.~(\ref{Schroedeqjump}). The weaker the interaction, the smaller can the jump rate $r$ be chosen, and the more efficient becomes the simulation. The proposed stochastic technique is expected to be particularly powerful when a problem is amenable to perturbation theory. By choosing a finite rate $r$, nonperturbative results can be obtained.

According to Eq.~(\ref{Hsimstate}), the action of the Hamiltonian $H^{(1)}$ in the jump (\ref{Schroedeqjump}) involves sums and integrals. In our stochastic approach, it is most natural to perform such sums and integrals by Monte Carlo methods. This introduces a second source of randomness into our Hilbert space process. After averaging, we still obtain a solution of the Schr\"odinger equation (\ref{Schroedeq}).

In $n$ jumps of the type (\ref{Schroedeqjump}), the trajectory typically progresses by $n$ times $1/r$. In view of the exponential time distribution, however, all positive values of time increments are possible, with the following probability density for the $n$-jump contribution at time $t$,
\begin{equation}\label{njumpprobdens}
    \int_0^t dt' r^n e^{-r( \tau_1 + \dots \tau_n )} \,
    \delta(\tau_1 + \dots \tau_n-t') \, e^{-r(t-t')} ,
\end{equation}
with the waiting times $\tau_1 \dots \tau_n$ for the jumps $1 \ldots n$ to occur. The last factor is the probability for having no further jump between $t'$ and $t$. In view of the $\delta$ function in Eq.~(\ref{njumpprobdens}), it is particular convenient to calculate the $n$-jump contribution to the Fourier transform (\ref{Romdef}) of the evolution operator. We then obtain
\begin{eqnarray}
    R_{\omega} &=& \sum_{n=0}^\infty (-1)^n (H^{(0)} - \omega - i\epsilon -ir)^{-1}
    \nonumber \\ &\times&
    \left[ (H^{(1)}+ir) (H^{(0)} - \omega - i\epsilon -ir)^{-1} \right]^n . \qquad
\label{integratedsim}
\end{eqnarray}
For $r=0$, this corresponds to the usual perturbation theory in Eq.~(\ref{Romperturb}), but the summed result is actually independent of $r$. So, we can eliminate the stochastic aspect of the jump rate in calculating Fourier transformed time correlation functions, but the randomness associated with Monte Carlo summations and integrations remains in the simulation. By varying the rate parameter $r$ in our simulations, we can go all the way from continuous integration (large $r$) to perturbation theory (small $r$). The simulation corresponds to the perturbation theory given by Eqs.~(\ref{Romperturb}) and (\ref{vacuumexpansion}). However, the formal regularization parameter $\epsilon$ is replaced by the rate parameter $r$, where $1/r$ is an average time step and $i r$ is added to the interaction. We have thus found a very simple recipe for how to proceed from perturbation theory to an integration scheme with average time step $1/r$.

\subsection{Procedure}
We are now in a position to describe the stochastic simulation procedure for calculating correlation functions. During the simulation of a trajectory in Hilbert space, we always have exactly one of the base vectors (\ref{simstate}) together with a corresponding complex coefficient. The free evolution between interactions is easy to evaluate because, during the entire simulation, we always keep eigenstates of the energy of the noninteracting system, as given in Eq.~(\ref{simstate}). Free evolution merely changes the coefficient but not the base vector.

Let us assume that we wish to calculate the time correlation function $\Dbra{0} A e^{-iHt} B \Dket{0}$ or its Fourier transform $\Dbra{0} A R_\omega B \Dket{0}$ occurring in Eqs.~(\ref{correlFT2}) or (\ref{stanfermprop}). The basic idea is to start from the vacuum state $\gf$ of the free theory, to initiate by a sufficiently large number of time steps, to apply the operator $B$, to perform the time steps for the evolution from $B$ to $A$, to apply the operator $A$, to finalize by a sufficiently large number of time steps, and to project onto $\gfa$. ``Initializing'' and ``finalizing'' should be recognized as the counterparts of the two parts of Eq.~(\ref{vacuumexpansion}); in a sense, we need to find the vacuum state $\gi$ of the interacting theory by ``equilibration'' when we start from the vacuum state of the free theory. For these equilibration steps, we should use Eq.~(\ref{integratedsim}) with $\omega = r = 0$. Going to the limit $r=0$, or a very large time step, is certainly appropriate for producing the vacuum state of the interacting theory. When we apply $R_\omega$ for the evolution from $B$ to $A$, the rate parameter $r$ and the number of interactions $n$ in Eq.~(\ref{integratedsim}) should be adjusted to the frequencies $\omega$ we would like to explore ($n \omega \sim r$).

Whenever we apply $H^{(1)}$ according to Eq.~(\ref{Hsimstate}), there is a number of summations and integrations to be carried out. For each sum or integral we choose only one single term; the full sums and integrals are obtained by averaging over many realizations. For example, we can choose the probability $p_{\rm rule}(j)$ for using one of the collision rules $j=1, \ldots 8$ to be $1/8$ and then multiply the estimate for the randomly chosen term $j$ by $8$ (more generally, by $1/p_{\rm rule}(j)$). The other summations and integrations in Eq.~(\ref{Hsimstate}) are performed with randomly selected particles, spins, polarizations or randomly selected momenta. Whenever we make a random choice, we need to introduce the reciprocal of the corresponding probability as a weight factor. In the final part of the simulation, one should only admit random selections that can still lead towards the free equilibrium state, with correspondingly larger probabilities and smaller weight factors.

A subtlety is introduced by the $\delta$ function in rule number seven in Eq.~(\ref{Hsimstate}), which reduces the number of integrations by one. As the rule is associated with the annihilation of a positron and we typically choose initial and final states that do not contain any positrons, we can always use the $\delta$ function to fix the momentum of the positron when it is created. As long as a positron exists in the simulation, we need to leave it open whether the positron can be randomly selected (if it is eventually annihilated by rule 3) or it is fixed (if it is eventually annihilated by rule 7).

Note that, in principle, the proper averages are obtained for any choice of the probabilities for randomly selecting rules, particles, spins, polarizations or momenta. However, the efficiency of the simulation, or the size of the statistical error bars, depends crucially on the choice of these probabilities for the simulation algorithm. The more uniform the weight factors, the more efficient is the simulation. The art of developing good Monte Carlo simulations consists in finding probabilities that reflect the summations and integrations in Eq.~(\ref{Hsimstate}) as faithfully as possible.

\subsection{Truncations}
In a practical implementation of the simulation procedure, a number of truncations are required. Most obviously, when the expression (\ref{integratedsim}) for $R_{\omega}$ is used, only a finite number of interactions can be kept (the number of interactions corresponds to the number of time steps). The smaller the frequency $\omega$, or the larger the probed time scale, the more interactions need to be kept. The required computer time grows linearly with the number of interactions.

It is convenient to limit the maximum number of electrons, positrons and photons present in a simulation at any given time. Limiting the number of simultaneously present field quanta corresponds to limiting the order of perturbation theory. Note however that, in the course of the simulation, a much larger number of electrons, positrons and photons can occur (and disappear). In that sense, we deal with a renormalized perturbation theory with interactions on all scales.

In practice, we actually set minimum and maximum values for the photon momenta to avoid possible infrared and ultraviolet problems. This is done through the choice of momenta according to a probability density $f(\bm{q})$. If one starts with a probability density for $q$, one needs to find an algorithm to sample according to that distribution. We hence prefer to start with an algorithm for generating $q$ and then determine the corresponding probability density. As we anticipate logarithmic divergences at large and small wave vectors in QED, a natural choice for the length of the wave vector of a photon is given by
\begin{equation}\label{qsampling}
    q = \left( \frac{q_{\rm max}}{q_{\rm min}} \right)^{\hat{r}} q_{\rm min}
    = q_{\rm max}^{\hat{r}} \, q_{\rm min}^{1-\hat{r}} ,
\end{equation}
where $q_{\rm min}$ and $q_{\rm max}$ are the lower and upper cutoffs and $\hat{r}$ is a uniform random number on the interval $[0,1]$. If the orientation of the vector $\bm{q}$ is chosen randomly, this choice corresponds to an isotropic power law probability density for the three-dimensional wave vector given by
\begin{equation}\label{qsamplingdistr}
    f(\bm{q}) = \frac{1}{4\pi \ln(q_{\rm max}/q_{\rm min})} \, \frac{1}{q^3} .
\end{equation}

\begin{figure}
\centerline{\epsfxsize=7.5cm \epsffile{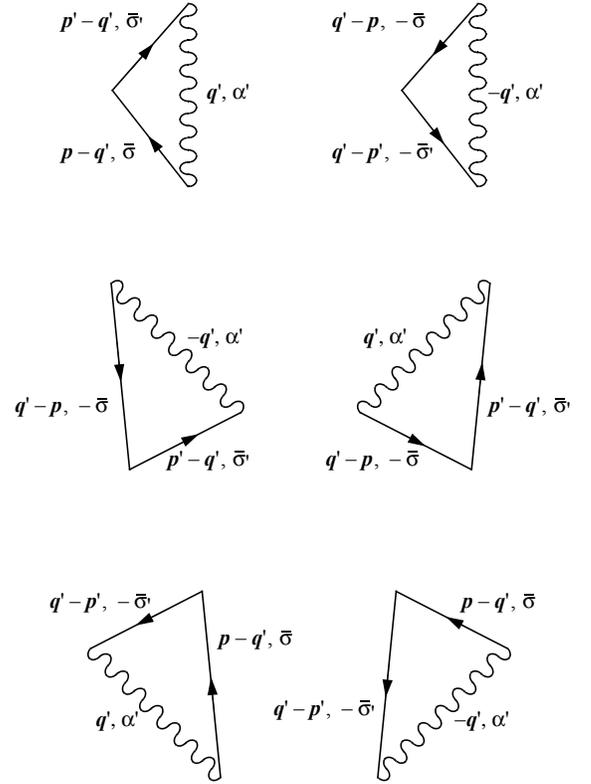}}
\caption[ ] {Feynman diagrams for the vertex function in the Hamiltonian approach; time increases from bottom to top.}
\label{qed_Feynman3D}
\end{figure}

As an alternative, one can try a uniform distribution of $q$,
\begin{equation}\label{qsamplingx}
    q = \hat{r} \, q_{\rm max} + (1-\hat{r}) \, q_{\rm min} ,
\end{equation}
which implies the probability density
\begin{equation}\label{qsamplingdistrx}
    f(\bm{q}) = \frac{1}{4\pi (q_{\rm max}-q_{\rm min})} \, \frac{1}{q^2} .
\end{equation}
The most efficient choice of the probability density $f(\bm{q})$ is expected to depend on the quantity of interest.

\subsection{Example}
\begin{figure}
\centerline{\epsfysize=11cm \epsffile{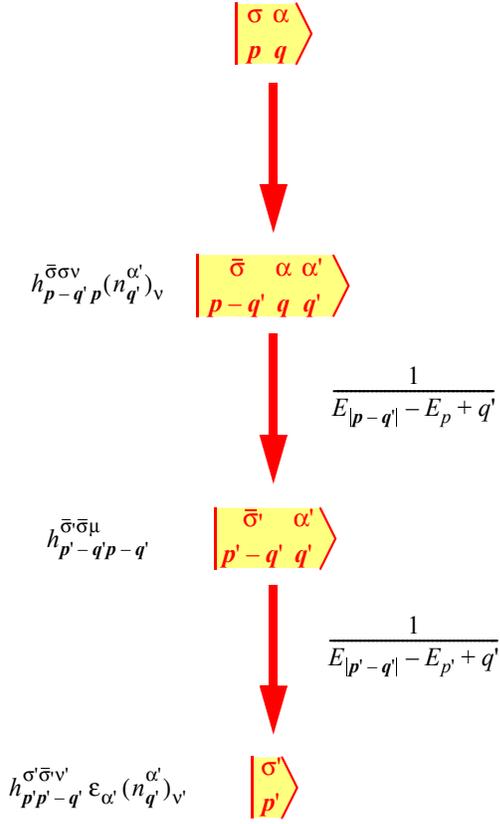}}
\caption[ ] {Sequence of interactions associated with the first Feynman diagram in Fig.~\ref{qed_Feynman3D}.}
\label{qed_sequence}
\end{figure}

In order to illustrate our simulation approach, we consider a very simple simulation in which the interaction is applied only three times. Even such a toy simulation contains interesting physics. According to Fig.~\ref{qed_Feynman}, such a simulation should be sufficient to get the leading-order correction to the effective collision amplitude and hence to the magnetic moment of the electron. The Feynman diagram in Fig.~\ref{qed_Feynman} represents the contribution to the vertex function given in Eq.~(\ref{vertexfct3}), that is, it involves momentum four-vectors and four-dimensional integrations. Our Hamiltonian time-evolution approach, on the other hand, is focused on time evolution, momentum three-vectors and three-dimensional integrations, as is clearly visible in all expressions for the various Hamiltonians. We need to identify the possible sequences of three interactions that are consistent with the Feynman diagram in Fig.~\ref{qed_Feynman} in our approach. If we amputate the external legs from this Feynman diagram and focus on the core triangle representing ${\Gamma^\mu}^{(3)}$, we obtain the Feynman diagrams for the corresponding three-step simulation shown in Fig.~\ref{qed_Feynman3D}. For our kinetic-theory-type simulation, these six diagrams are not equivalent because they correspond to the six possible permutations of three interaction events occurring in the course of time; they correspond to the six terms of Eq.~(\ref{correlFT3}), in the same order, from which we get the respective frequencies and signs coming with each term (possible minus signs are actually taken into account by the amputation procedure so that we can use positive signs for all terms contributing to the vertex function). The information about the full diagrams with external legs can be reconstructed by extending the fermion line by an incoming electron of momentum $\bm{p}$ and spin $\sigma$ and an outgoing electron of momentum $\bm{p}'$ and spin $\sigma'$, and by attaching an incoming photon line with momentum $\bm{q}=\bm{p}'-\bm{p}$ and spacetime index $\mu$ to the kink of the fermion line. Therefore, the six diagrams in Fig.~\ref{qed_Feynman3D} specify the three-step simulations for the vertex function in a unique way.

\begin{figure}
\centerline{\epsfysize=11cm \epsffile{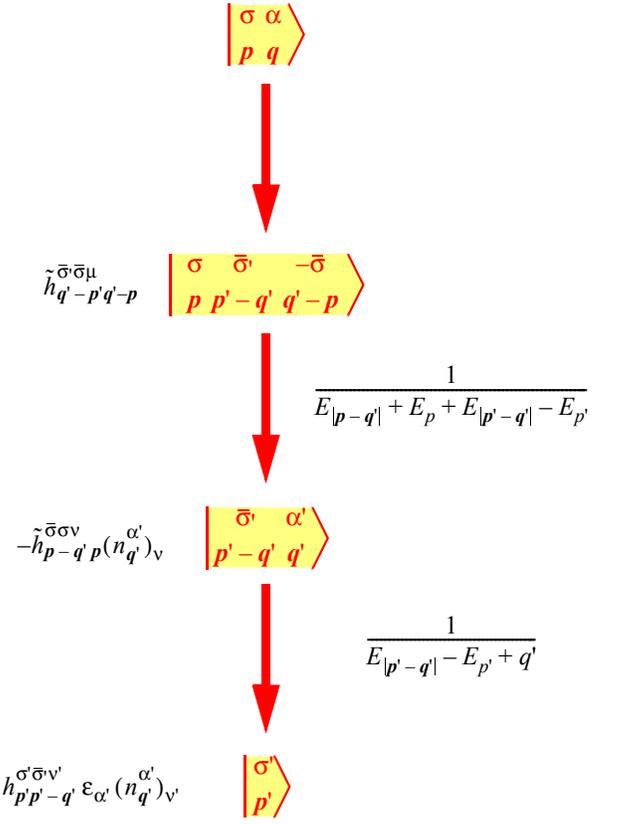}}
\caption[ ] {Sequence of interactions associated with the fourth Feynman diagram in Fig.~\ref{qed_Feynman3D}.}
\label{qed_sequencex}
\end{figure}

The first two Feynman diagrams in Fig.~\ref{qed_Feynman3D} correspond to absorption of an external photon by an electron or positron, the next two diagrams correspond to pair production, and the last two diagrams correspond to pair annihilation. The simulation steps associated with the first diagram are shown in detail in Fig.~\ref{qed_sequence}. In the middle, we show the sequence of states of the form (\ref{simstate}) for the trajectory in Fock space; for this diagram, the visited states consist of only one electron and of up to two photons. On the left-hand side, we show the collision amplitudes occurring for each of the three interactions according to the collision rules in Eq.~(\ref{Hsimstate}). For the contribution associated with the kink in the fermion line, the polarization vector is missing because, in the definition (\ref{hphotemisseff}) of the effective collision amplitude, we have multiplied $\Gamma^\mu$ with the spinors but not with the polarization vectors. On the right-hand side of Fig.~\ref{qed_sequence}, we show the factors resulting from the free evolution according to $R^{(0)}$ with the frequencies obtained from Eq.~(\ref{correlFT3}) where, in view of Eq.~(\ref{formfacintro}), we choose $\omega=E_p$ and $\omega'=E_{p'}$ (Eq.~(\ref{deffvertex}) then implies that the incoming and outgoing electrons are on the mass shell). To amputate the external legs, only the propagating particles shown in Fig.~\ref{qed_Feynman3D} need to be taken into account for the energy calculations.

For example, in the first step of Fig.~\ref{qed_sequence}, an electron with momentum $\bm{p}$ and spin $\sigma$ emits a photon with momentum $\bm{q}'$ and polarization $\alpha'$, whereupon the electron momentum changes to $\bm{p}-\bm{q}'$ and the spin to $\bar{\sigma}$. From the fourth rule in Eq.~(\ref{Hsimstate}) we hence read of the collision amplitude
$h^{\bar{\sigma} \sigma \nu}_{\bm{p}-\bm{q}' \, \bm{p}} \, ( n^{\alpha'}_{\bm{q}'} )_\nu$, which is the first factor on the left-hand side of Fig.~\ref{qed_sequence}. Of the electron and the two photons of the resulting state, only the electron (with energy $E_{|\bm{p}-\bm{q}'|}$) and one photon (with energy $q'$) are propagated in the lower half of the first Feynman diagram in Fig.~\ref{qed_Feynman3D} because the external photon (with energy $q$) is amputated. For the action of the free evolution operator, this implies
\begin{equation}\label{R0facexample}
    R^{(0)}_\omega = \frac{1}{E_{|\bm{p}-\bm{q}'|}+q'-\omega} ,
\end{equation}
where the frequency $\omega$ has been identified from the first term of Eq.~(\ref{correlFT3}), which corresponds to the first Feynman diagram in Fig.~\ref{qed_Feynman3D}. With $\omega=E_p$, we obtain the first factor on the right-hand side of Fig.~\ref{qed_sequence}. The remaining factors can be found by analogous arguments.

A sequence of steps like the one shown in Fig.~\ref{qed_sequence} can be associated with each of the six Feynman diagrams in Fig.~\ref{qed_Feynman3D} (a further example is given in Fig.~\ref{qed_sequencex}). By collecting all the factors associated with each of the six diagrams and using Eq.~(\ref{modcomplete}), we obtain the following effective collision amplitude for a three-step simulation (the order of the terms corresponds to the order of the Feynman diagrams in Fig.~\ref{qed_Feynman3D} and hence also to the order of the terms in Eq.~(\ref{correlFT3})):

\begin{widetext}

\begin{eqnarray}
    {\cal H}^{\sigma' \sigma \mu}_{\bm{p}' \bm{p}} &=& h^{\sigma' \sigma \mu}_{\bm{p}' \bm{p}} +
    \int d^3q' \frac{1}{E_{|\bm{p}-\bm{q}'|}-E_p+q'} \frac{1}{E_{|\bm{p}'-\bm{q}'|}-E_{p'}+q'} \,
    h^{\sigma' \bar{\sigma}' \nu'}_{\bm{p}' \, \bm{p}'-\bm{q}'} \,
    h^{\bar{\sigma}' \bar{\sigma} \mu}_{\bm{p}'-\bm{q}' \, \bm{p}-\bm{q}'} \,
    h^{\bar{\sigma} \sigma \nu}_{\bm{p}-\bm{q}' \, \bm{p}} \, \eta_{\nu \nu'}
    \nonumber \\
    &+& \int d^3q' \frac{1}{E_{|\bm{p}-\bm{q}'|}+E_p+q'} \frac{1}{E_{|\bm{p}'-\bm{q}'|}+E_{p'}+q'} \,
    \tilde{h}^{\sigma' \bar{\sigma}' \nu'}_{-\bm{p}' \,\bm{q}'-\bm{p}'} \,
    h^{\bar{\sigma}' \bar{\sigma} \mu}_{\bm{q}'-\bm{p}' \, \bm{q}'-\bm{p}} \,
    \tilde{h}^{\bar{\sigma} \sigma \nu}_{\bm{p}-\bm{q}' \, \bm{p}} \, \eta_{\nu \nu'}
    \nonumber \\
    && \hspace{-4.5em} - \, \int d^3q' \frac{1}{E_{|\bm{p}-\bm{q}'|}+E_p+E_{|\bm{p}'-\bm{q}'|}-E_{p'}}
    \left( \frac{1}{E_{|\bm{p}-\bm{q}'|}+E_p+q'} + \frac{1}{E_{|\bm{p}'-\bm{q}'|}-E_{p'}+q'} \right) \,
    h^{\sigma' \bar{\sigma}' \nu'}_{\bm{p}' \, \bm{p}'-\bm{q}'} \,
    \tilde{h}^{\bar{\sigma}' \bar{\sigma} \mu}_{\bm{q}'-\bm{p}' \, \bm{q}'-\bm{p}} \,
    \tilde{h}^{\bar{\sigma} \sigma \nu}_{\bm{p}-\bm{q}' \, \bm{p}} \, \eta_{\nu \nu'}
    \nonumber \\
    && \hspace{-4.5em} - \, \int d^3q' \frac{1}{E_{|\bm{p}-\bm{q}'|}-E_p+E_{|\bm{p}'-\bm{q}'|}+E_{p'}}
    \left( \frac{1}{E_{|\bm{p}-\bm{q}'|}-E_p+q'} + \frac{1}{E_{|\bm{p}'-\bm{q}'|}+E_{p'}+q'} \right) \,
    \tilde{h}^{\sigma' \bar{\sigma}' \nu'}_{-\bm{p}' \,\bm{q}'-\bm{p}'} \,
    \tilde{h}^{\bar{\sigma}' \bar{\sigma} \mu}_{\bm{p}'-\bm{q}' \, \bm{p}-\bm{q}'} \,
    h^{\bar{\sigma} \sigma \nu}_{\bm{p}-\bm{q}' \, \bm{p}} \, \eta_{\nu \nu'} \,.
    \nonumber \\ &&
\label{heffectiveZee}
\end{eqnarray}

\end{widetext}

With the exception of photon emission by a positron, all possible collision rules have been used to obtain Eq.~(\ref{heffectiveZee}), so that our toy simulation is a serious test of the proposed simulation approach. For a formal proof of the suitability of the amputation rules in our simulation, one should compare Eqs.~(\ref{vertexfct33}) and (\ref{heffectiveZee}). After multiplying Eq.~(\ref{vertexfct33}) with the proper factors to obtain the effective collision amplitude (\ref{hphotemisseff}), we see the claimed equivalence after using the definitions (\ref{projectorele}) and (\ref{projectorpos}) for the electron and positron projectors. This reformulation demonstrates nicely that perturbation theory focuses on propagators whereas our simulations focus on collision rules and collision amplitudes.

For our actual simulations of the various terms in Eq.~(\ref{heffectiveZee}), we choose the convenient momentum vectors
\begin{equation}\label{ps4actsim}
    \bm{p} = \left( \begin{array}{c}
       y \\
       -x/2 \\
       0 \\
    \end{array} \right) \,, \qquad
    \bm{p}' = \left( \begin{array}{c}
       y \\
       x/2 \\
       0 \\
    \end{array} \right) \,,
\end{equation}
and we hence have
\begin{equation}\label{energy4actsim}
    E_p^2 = E_{p'}^2 = \frac{x^2}{4} + y^2 + m^2 .
\end{equation}
The corresponding polarization vectors $n^\alpha_{\bm{q}}$ defined in Eqs.~(\ref{polarization0})--(\ref{polarization3}) for $\alpha=0,1,2,3$ are given by
\begin{equation}\label{pols4actsim}
    \left( \begin{array}{c}
       1 \\
       0 \\
       0 \\
       0 \\
    \end{array} \right) \,, \quad
    \left( \begin{array}{c}
       0 \\
       1 \\
       0 \\
       0 \\
    \end{array} \right) \,, \quad
    \left( \begin{array}{c}
       0 \\
       0 \\
       0 \\
       -1 \\
    \end{array} \right) \,, \quad
    \left( \begin{array}{c}
       0 \\
       0 \\
       1 \\
       0 \\
    \end{array} \right) \,,
\end{equation}
respectively.

To identify the form factors $F$ and $G$, we have listed the two building blocks occurring in the representation (\ref{formfacintro}) for our choice (\ref{ps4actsim}) of the momenta $\bm{p}$ and $\bm{p}'$ in Tables \ref{tableGordon1} and \ref{tableGordon2}. According to Eq.~(\ref{pols4actsim}), the rows for the spatial components $\mu=1$ and $\mu=3$ correspond to the physically observable transverse photon polarizations. For symmetry reasons, we can focus on the two double-underlined components in these tables.

\begin{table}
\caption[ ]{Components of the four-vector $2m \, \bar{u}^{\sigma'}_{\bm{p}'} \, \gamma^\mu \, u^{\sigma}_{\bm{p}}$ for the momenta given in Eq.~(\ref{ps4actsim}); the pairs of spin values $\sigma,\sigma'$ are given in the first row, the values of the spacetime index $\mu$ in the first column; our data analysis is focused on the two double-underlined components}
\begin{tabular}{|c|c|c|c|c|}
\hline
& $\frac{1}{2}$, $\frac{1}{2}$ & $-\frac{1}{2}$, $-\frac{1}{2}$ & $\frac{1}{2}$, $-\frac{1}{2}$ & $-\frac{1}{2}$, $\frac{1}{2}$ \\
\hline
$0$ &
$2m+\frac{2y^2-xyi}{E_p+m}$ & $2m+\frac{2y^2+xyi}{E_p+m}$ & $0$ & $0$ \\
$1$ &
$\underline{\underline{2y-ix}}$ & $2y+ix$ & $0$ & $0$ \\
$2$ &
$0$ & $0$ & $0$ & $0$ \\
$3$ &
$0$ & $0$ & $ix$ & $\underline{\underline{ix}}$ \\
\hline
\end{tabular}
\label{tableGordon1}
\end{table}

\begin{table}
\caption[ ]{Components of the four-vector $\bar{u}^{\sigma'}_{\bm{p}'} \, u^{\sigma}_{\bm{p}} \, (p+p')^\mu$ for the momenta given in Eq.~(\ref{ps4actsim}); the pairs of spin values $\sigma,\sigma'$ are given in the first row, the values of the spacetime index $\mu$ in the first column; our data analysis is focused on the two double-underlined components}
\begin{tabular}{|c|c|c|c|c|}
\hline
& $\frac{1}{2}$, $\frac{1}{2}$ & $-\frac{1}{2}$, $-\frac{1}{2}$ & $\frac{1}{2}$, $-\frac{1}{2}$ & $-\frac{1}{2}$, $\frac{1}{2}$ \\
\hline
$0$ &
$E_p\left[2+\frac{x^2+2xyi}{2m(E_p+m)}\right]$ & $E_p\left[2+\frac{x^2-2xyi}{2m(E_p+m)}\right]$ & $0$ & $0$ \\
$1$ &
$\underline{\underline{y\left[2+\frac{x^2+2xyi}{2m(E_p+m)}\right]}}$ & $y\left[2+\frac{x^2-2xyi}{2m(E_p+m)}\right]$ & $0$ & $0$ \\
$2$ &
$0$ & $0$ & $0$ & $0$ \\
$3$ &
$0$ & $0$ & $0$ & $\underline{\underline{0}}$ \\
\hline
\end{tabular}
\label{tableGordon2}
\end{table}

From Table~\ref{tableGordon1}, we conclude that the imaginary part of
${\cal H}^{1/2, 1/2, 1}_{\bm{p}' \bm{p}} + {\cal H}^{1/2, -1/2, 3}_{\bm{p}' \bm{p}}$
results exclusively from the second contribution in Eq.~(\ref{formfacintro}). From Eqs.~(\ref{formfacintro}), (\ref{hphotemisseff}) and Table~\ref{tableGordon2}, we hence find the form factor
\begin{eqnarray}
    G(x) &=& \frac{\sqrt{(4\pi)^3} \, m E_p (E_p+m)}{e_0 y^2 \sqrt{x}}
    \nonumber \\ &\times &
    {\rm Im} \left( {\cal H}^{1/2, 1/2, 1}_{\bm{p}' \bm{p}}
    + {\cal H}^{1/2, -1/2, 3}_{\bm{p}' \bm{p}} \right).
\label{Gfromsim}
\end{eqnarray}
By extrapolating to $x=0$, we then obtain the $g$-factor for the electron according to Eq.~(\ref{gfacelectron}).

\begin{table}
\caption[ ]{Components of the four-vector $i \, \bar{u}^{\sigma'}_{\bm{p}'} \, \sigma^{\mu\nu} (p'-p)^\nu \, u^{\sigma}_{\bm{p}}$ for the momenta given in Eq.~(\ref{ps4actsim}); the pairs of spin values $\sigma,\sigma'$ are given in the first row, the values of the spacetime index $\mu$ in the first column; our data analysis is focused on the two double-underlined components}
\begin{tabular}{|c|c|c|c|c|}
\hline
& $\frac{1}{2}$, $\frac{1}{2}$ & $-\frac{1}{2}$, $-\frac{1}{2}$ & $\frac{1}{2}$, $-\frac{1}{2}$ & $-\frac{1}{2}$, $\frac{1}{2}$ \\
\hline
$0$ &
$\frac{x^2+2xyi}{2m}$ & $\frac{x^2-2xyi}{2m}$ & $0$ & $0$ \\
$1$ &
$\underline{\underline{ix+y\frac{x^2+2xyi}{2m(E_p+m)}}}$ & $-ix+y\frac{x^2-2xyi}{2m(E_p+m)}$ & $0$ & $0$ \\
$2$ &
$0$ & $0$ & $0$ & $0$ \\
$3$ &
$0$ & $0$ & $-ix$ & $\underline{\underline{-ix}}$ \\
\hline
\end{tabular}
\label{tableGordon3}
\end{table}

To gain a better intuition why the deviation of the $g$-factor from $2$ is given by $G(0)$, one should look at Table~\ref{tableGordon3}. The observation ``Table~\ref{tableGordon1}$\,=\,$Table~\ref{tableGordon2}$\,-\,$Table~\ref{tableGordon3}'' is known as the Gordon decomposition. By means of this decomposition, one can rewrite the representation (\ref{formfacintro}) of the vertex functions in terms of the momentum- and spin-dependent terms in Tables~\ref{tableGordon2} and \ref{tableGordon3}, so that the occurrence of the magnetic moment becomes plausible.

For our toy simulation of ${\cal H}^{\sigma' \sigma \mu}_{\bm{p}' \bm{p}}$, the summations over spin values and spatial components in Eq.~(\ref{heffectiveZee}) can actually be performed in a deterministic manner rather than by Monte Carlo methods. Only the integrations over the photon momentum $\bm{q}'$ in Eq.~(\ref{heffectiveZee}) need to be done by Monte Carlo simulations. We have tried the sampling procedures (\ref{qsampling}) and (\ref{qsamplingx}) for the photon momentum; it turned out that, for our simulation of the form factor $G$, the error bars for the procedure (\ref{qsampling}) are about two times larger than for the procedure (\ref{qsamplingx}), so that we adopted the latter one. To reproduce the symmetry properties in Tables~\ref{tableGordon1}--\ref{tableGordon3} rigorously, we have actually symmetrized the integrands in Eq.~(\ref{heffectiveZee}) in the two-component of $\bm{q}'$. All quantities have been made dimensionless by setting $m=1$. The full simulation code consists of about a hundred commands.

\begin{figure}
\centerline{\epsfxsize=8.5cm \epsffile{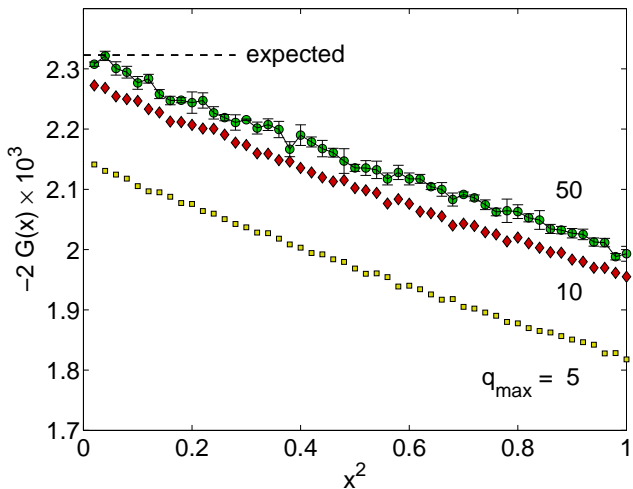}}
\caption[ ] {Simulation results for the form factor $G(x)$ of the electron.}
\label{qed_gresults}
\end{figure}

\begin{figure}
\centerline{\epsfxsize=8.5cm \epsffile{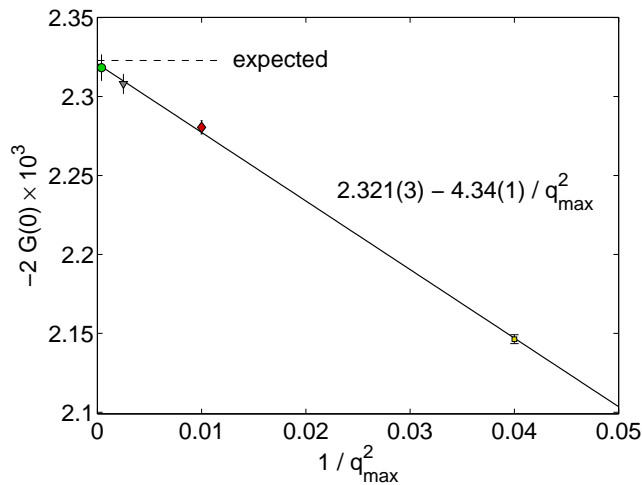}}
\caption[ ] {Extrapolation of the simulation results for the form factor $G(0)$ of the electron given in Table~\ref{tablecutoffG} to infinite cutoff; the dashed line represents a linear fit to the data points.}
\label{qed_gextrapol}
\end{figure}

Our simulation results for the form factor of the electron for $y=2$ are shown in Figure~\ref{qed_gresults}. According to Eq.~(\ref{Gfromsim}), the result for $G(x)$ should be independent of our choice of $y$; an intermediate value of $y$ seems to be most natural for the simulations (we have checked that $y=5$ works equally well). For each data point, $40$ million photon momenta have been sampled for the integrations. The required computer time on a single Intel Xeon $2.6$ GHz processor is about two minutes per data point. Curves for $q_{\rm min}=0$ and three different values of $q_{\rm max}$ are shown Figure~\ref{qed_gresults}. For small $x$, each of the curves is linear in $x^2$ and can easily be extrapolated to $x=0$. The extrapolated results are collected in Table~\ref{tablecutoffG}. Larger $q_{\rm max}$ leads to larger error bars. One should hence keep the cutoff as small as possible and then extrapolate to infinite cutoff, as shown in Figure~\ref{qed_gextrapol}. Our final extrapolated simulation result for infinite cutoff corresponds to
\begin{equation}\label{gfacelectronsim}
    g = 2.002321(3) \, ,
\end{equation}
in perfect agreement with Eq.~(\ref{gfacelectron1}). Of course, we could easily achieve higher precision by investing more than two minutes of CPU time per data point. However, the main point of our toy simulation is to demonstrate that a short and simple simulation program can reproduce the anomalous magnetic moment of the electron with minor computational effort. In view of our unsophisticated Monte Carlo sampling procedure, the efficiency of the simulations is truly amazing.

\begin{table}
\caption[ ]{Extrapolated simulation results for the form factor $G(0)$ of the electron for various values of the cutoff $q_{\rm max}$}
\begin{tabular}{|c|c|}
\hline
$q_{\rm max}$ & $-2G(0) \times 10^3$ \\
\hline
$5$  & $2.146(2)$ \\
$10$ & $2.281(4)$ \\
$20$ & $2.308(6)$ \\
$50$ & $2.318(8)$ \\
\hline
\end{tabular}
\label{tablecutoffG}
\end{table}

With the help of symbolic computation (series expansions and integrations handled by Mathematica), the four integrals in Eq.~(\ref{heffectiveZee}) in the limit of small $x$ and $y$ have actually been calculated in closed form by Martin Kr\"oger \cite{mkprivate}, thus reproducing Schwinger's famous result for $g$ given in Eq.~(\ref{gfacelectron1}). The resulting relative contributions to $g-2$ of the four integrals in Eq.~(\ref{heffectiveZee}) are given by $( 47 + 7 \pi + 34 \ln 2 ) / 30 \approx 3.085$ for absorption of a photon by an electron, $( -21 - 7 \pi + 34 \ln 2) / 30  \approx -0.647$ for absorption of a photon by a positron, and equal values of $( 1 - 17 \ln 2 ) / 15 \approx -0.719$ for pair production and pair annihilation. Simulations show that, for increasing $y$, the contributions from absorption of a photon by a positron, pair production and pair annihilation become less relevant (while the total result, apart from the magnitude of the error bars, is independent of $y$, as pointed out before). For large $y$, say $y \geq 100$, the result for $g$ is entirely dominated by the absorption of a photon by an electron, that is, by the first Feynman diagram in Fig.~\ref{qed_Feynman3D} or, in other words, by the sequence of collisions shown in Fig.~\ref{qed_sequence}.

\section{Summary and discussion}\label{secsd}
We have rewritten QED as a kinetic-theory-like evolution equation with complex collision amplitudes in the Fock space of electrons, positrons and photons. By using longitudinal and temporal in addition to transverse photons, the collision rules for photon emission/absorption and electron-positron pair production/annihilation become particularly simple. This specific procedure of introducing unphysical degrees of freedom corresponds to choosing a gauge. The Fourier transforms of time-ordered correlation functions can then be evaluated by stochastic simulations of trajectories in Fock space consisting of sequences of such collision processes. Summations and integrations introduced by the interaction Hamiltonian are evaluated by Monte Carlo techniques. In the limit of the average time step of the simulations going to infinity, we recover perturbation theory. Even for fundamental theoretical developments, the kinetic-theory-like formulation of QED with four types of photons may be advantageous.

An illustrative toy simulation used here to obtain the leading-order contribution to the anomalous magnetic moment of the electron is very close to perturbation theory, which is known as an extremely successful tool in QED. However, the purpose of the proposed stochastic particle simulations is not to obtain high-precision results or to compete with perturbation theory. The goal rather is to provide a flexible tool that can go beyond perturbation theory (by choosing a finite average time step).

In more typical simulations with many time finite steps one can, for example, study propagators for different values of the ultraviolet cutoff $q_{\rm max}$ and track how the (bare) mass and charge need to be changed with $q_{\rm max}$ to get consistent results for the frequency dependence of the propagator. With such a procedure, one can extract both the physical properties of the propagators and the renormalization-group flow of the model parameters (mass and charge). Additional physical properties, such as form factors, can also be obtained by such simulations.

Even in QED there are questions of fundamental importance that cannot be conclusively answered by perturbation theory. Most important is the question whether the bare charge diverges at a finite ultraviolet cutoff, at the so-called Landau pole. Such a pole would intrinsically indicate a limitation of QED at high energies and would imply a vanishing coupling constant in the absence of an ultraviolet cutoff (that is, triviality of pure QED). This question has been addressed by extensive lattice simulations of QED (see, for example, \cite{KogutDagoKoc88,KogutDagoKoc89,KimKogutLomb02,Gockeleretal90,Gockeleretal92,Gockeleretal97,Gockeleretal98}), which reveal a second-order chiral phase transition with a nonperturbative critical point in the strong-coupling regime (which does not correspond to a continuum version of QED). The lattice simulations imply that QED is a valid theory only for small renormalized charges \cite{Gockeleretal92}. Full consistency of QED can only be obtained in the bigger settings of electroweak interactions or the standard model. In view of these severe limitations of pure QED, it will be interesting to verify the consistency of the insights obtained from field- and particle-based simulations in the low-energy regime. In particular, one should compare the renormalization-group flow of the bare mass and charge with the ultraviolet cutoff. A major advantage of our particle-based simulations is that significantly larger variations of the cutoff can be achieved than with four-dimensional lattice simulations.

A regularization mechanism through an ultraviolet cutoff in momentum space is significantly different from a lattice formulation. Consistent simulation results in the low-energy domain of QED would hence be encouraging; although they cannot put QED on solid ground as a stand-alone quantum field theory, they would explain the enormous success of perturbative QED. For our stochastic simulation technique, still another regularization mechanism would be very natural: a rapid dissipative smoothing of the fields on short length scales \cite{hco200}. Such a dynamic coarse-graining approach to quantum field theory actually provided the original motivation for kinetic-theory-like stochastic simulation technique developed in the present work.

A generalization of the proposed simulation techniques from QED to nonabelian gauge theories would clearly be desirable. For quantum chromodynamics (QCD), a naive generalization of the ideas of Gupta \cite{Gupta50} and Bleuler \cite{Bleuler50} would suggest to introduce longitudinal and temporal gluons leading to particularly simple gluon propagators. However, additional ghost particles related to the Fadeev-Popov method for treating nonabelian gauge theories in the path-integral approach are required. The generalization of the ideas of Gupta and Bleuler to obtain a manifestly covariant canonical quantization of Yang-Mills theories has actually been developed in a series of papers by Kugo and Ojima \cite{KugoOjima78a,KugoOjima78b,KugoOjima79a,KugoOjima79b}. As a key ingredient to their approach, Kugo and Ojima use the conserved charges generating BRST transformations \cite{BecchiRouetStora76,Tyutin75} to characterize the physical states. In addition to the ghost particles, three- and four-gluon collisions make the kinetic theory for QCD more complicated than for QED. For electroweak interactions, an additional complication arises: the Higgs particle needs to be included into the kinetic-theory-like description. Although the details are considerably more complicated, the ideas of the present paper can be generalized to simulate the Yang-Mills theories for QCD and electroweak interactions and hence all parts of the standard model.

\begin{acknowledgments}
I am indebted to Martin Kr\"oger for supporting the final steps of this development by symbolic computations and by performing the actual simulations.\\[5mm]
\end{acknowledgments}

\appendix

\section*{Dirac matrices and spinors}\label{Diracnotation}
Pauli's famous $2 \times 2$ spin matrices,
\begin{equation}\label{Paulimatrices}
    \sigma^1 = \left( \begin{array}{rr}
      0 & 1 \\
      1 & 0 \\
    \end{array} \right) , \quad
    \sigma^2 = \left( \begin{array}{rr}
      0 & -i \\
      i & 0 \\
    \end{array} \right) , \quad
    \sigma^3 = \left( \begin{array}{rr}
      1 & 0 \\
      0 & -1 \\
    \end{array} \right) ,
\end{equation}
can be doubled to act on the upper (electron) and lower (positron) halves of spinors by defining the corresponding $4 \times 4$ matrices
\begin{equation}\label{Paulimatrices4}
    \sigma^{jk} = \epsilon^{jkl} \left( \begin{array}{cc}
      \sigma^l & 0 \\
      0 & \sigma^l \\
    \end{array} \right)  \quad \mbox{for } j,k=1,2,3 ,
\end{equation}
along with $\sigma^{00} = 0$ and $\sigma^{0j} = - \sigma^{j0} = i \, \gamma^0 \gamma^j$, where Dirac's $4 \times 4$ matrices $\gamma^j$ are also defined in terms of Pauli's matrices,
\begin{equation}\label{Diracmatrices}
    \gamma^j = \left( \begin{array}{cc}
      0 & \sigma^j \\
      -\sigma^j & 0 \\
    \end{array} \right) \quad \mbox{for } j=1,2,3 ,
\end{equation}
and we have further used the definition
\begin{equation}\label{Diracmatrices0}
    \gamma^0 = \left( \begin{array}{rrrr}
      1 & 0 & 0 & 0 \\
      0 & 1 & 0 & 0 \\
      0 & 0 & -1 & 0 \\
      0 & 0 & 0 & -1 \\
    \end{array} \right) \,.
\end{equation}
With these definitions of Dirac's matrices $\gamma^\mu$ and the definitions of spinors in Eqs.~(\ref{spinorsu1})--(\ref{spinorsv2}), one can evaluate the products listed in Tables~\ref{tablecouplmatrix} and \ref{tablecouplmatriy}.

\begin{widetext}

\begin{table}
\caption[ ]{Components of $2m \, \bar{u}^{\sigma}_{\bm{p}} \, \gamma^\mu \, u^{\sigma'}_{\bm{p}'}/ \sqrt{(E_p+m)(E_{p'}+m)}$ in terms of $\hat{p}_j = p_j/(E_p+m)$, $\hat{p}_\pm = \hat{p}_1 \pm i\hat{p}_2$, $\hat{p}'_j = p'_j/(E_{p'}+m)$ and $\hat{p}'_\pm = \hat{p}'_1 \pm i\hat{p}'_2$; the pairs of spin values $\sigma,\sigma'$ are given in the first row, the values of the spacetime index $\mu$ in the first column}
\begin{tabular}{|c|c|c|c|c|}
\hline
& $1/2$, $1/2$ & $-1/2$, $-1/2$ & $1/2$, $-1/2$ & $-1/2$, $1/2$ \\
\hline
$0$ &
$1+\hat{p}_-\hat{p}'_++\hat{p}_3\hat{p}'_3$ &
$1+\hat{p}_+\hat{p}'_-+\hat{p}_3\hat{p}'_3$ &
$\hat{p}_3\hat{p}'_--\hat{p}_-\hat{p}'_3$ &
$-\hat{p}_3\hat{p}'_++\hat{p}_+\hat{p}'_3$ \\
$1$ &
$\hat{p}_-+\hat{p}'_+$ &
$\hat{p}_++\hat{p}'_-$ &
$\hat{p}_3-\hat{p}'_3$ &
$-\hat{p}_3+\hat{p}'_3$ \\
$2$ &
$i\hat{p}_--i\hat{p}'_+$ &
$-i\hat{p}_++i\hat{p}'_-$ &
$-i\hat{p}_3+i\hat{p}'_3$ &
$-i\hat{p}_3+i\hat{p}'_3$ \\
$3$ &
$\hat{p}_3+\hat{p}'_3$ &
$\hat{p}_3+\hat{p}'_3$ &
$-\hat{p}_-+\hat{p}'_-$ &
$\hat{p}_+-\hat{p}'_+$ \\
\hline
\end{tabular}
\label{tablecouplmatrix}
\end{table}

\begin{table}
\caption[ ]{Components of $2m \, \bar{v}^{-\sigma}_{-\bm{p}} \, \gamma^\mu \, u^{\sigma'}_{\bm{p}'}/ \sqrt{(E_p+m)(E_{p'}+m)}$ in terms of $\hat{p}_j = p_j/(E_p+m)$, $\hat{p}_\pm = \hat{p}_1 \pm i\hat{p}_2$, $\hat{p}'_j = p'_j/(E_{p'}+m)$ and $\hat{p}'_\pm = \hat{p}'_1 \pm i\hat{p}'_2$; the pairs of spin values $\sigma,\sigma'$ are given in the first row, the values of the spacetime index $\mu$ in the first column}
\hspace*{-3em}
\begin{tabular}{|c|c|c|c|c|}
\hline
& $1/2$, $1/2$ & $-1/2$, $-1/2$ & $1/2$, $-1/2$ & $-1/2$, $1/2$ \\
\hline
$0$ &
$-\hat{p}_3+\hat{p}'_3$ &
$\hat{p}_3-\hat{p}'_3$ &
$-\hat{p}_-+\hat{p}'_-$ &
$-\hat{p}_++\hat{p}'_+$ \\
$1$ &
$-\hat{p}_3\hat{p}'_+-\hat{p}_-\hat{p}'_3$ &
$\hat{p}_3\hat{p}'_-+\hat{p}_+\hat{p}'_3$ &
$1-\hat{p}_-\hat{p}'_-+\hat{p}_3\hat{p}'_3$ &
$1-\hat{p}_+\hat{p}'_++\hat{p}_3\hat{p}'_3$ \\
$2$ &
$i\hat{p}_3\hat{p}'_+-i\hat{p}_-\hat{p}'_3$ &
$i\hat{p}_3\hat{p}'_--i\hat{p}_+\hat{p}'_3$ &
$-i(1+\hat{p}_-\hat{p}'_-+\hat{p}_3\hat{p}'_3)$ &
$i(1+\hat{p}_+\hat{p}'_++\hat{p}_3\hat{p}'_3)$ \\
$3$ &
$1+\hat{p}_-\hat{p}'_+-\hat{p}_3\hat{p}'_3$ &
$-1-\hat{p}_+\hat{p}'_-+\hat{p}_3\hat{p}'_3$ &
$-\hat{p}_3\hat{p}'_--\hat{p}_-\hat{p}'_3$ &
$-\hat{p}_3\hat{p}'_+-\hat{p}_+\hat{p}'_3$ \\
\hline
\end{tabular}
\label{tablecouplmatriy}
\end{table}

\end{widetext}

As the spinors $v$ are obtained from the spinors $u$ by exchanging their upper and lower halves and flipping the spins, we immediately obtain the relations
\begin{equation}\label{uuvvident}
    \bar{u}^{\sigma}_{\bm{p}} \, \gamma^\mu \, u^{\sigma'}_{\bm{p}'} =
    \bar{v}^{-\sigma}_{\bm{p}} \, \gamma^\mu \, v^{-\sigma'}_{\bm{p}'} ,
\end{equation}
and similarly
\begin{equation}\label{uvvuident}
    \bar{u}^{\sigma}_{\bm{p}} \, \gamma^\mu \, v^{\sigma'}_{\bm{p}'} =
    \bar{v}^{-\sigma}_{\bm{p}} \, \gamma^\mu \, u^{-\sigma'}_{\bm{p}'} .
\end{equation}
By inspection of Table~\ref{tablecouplmatrix}, one can conveniently verify the symmetry property
\begin{equation}\label{uuident}
    \bar{u}^{\sigma}_{\bm{p}} \, \gamma^\mu \, u^{\sigma'}_{\bm{p}'} =
    (\bar{u}^{\sigma'}_{\bm{p}'} \, \gamma^\mu \, u^{\sigma}_{\bm{p}})^* ,
\end{equation}
and, in view of Eq.~(\ref{uuvvident}), we similarly have
\begin{equation}\label{vvident}
    \bar{v}^{\sigma}_{\bm{p}} \, \gamma^\mu \, v^{\sigma'}_{\bm{p}'} =
    (\bar{v}^{\sigma'}_{\bm{p}'} \, \gamma^\mu \, v^{\sigma}_{\bm{p}})^* .
\end{equation}
These symmetry properties, as well as the further relation
\begin{equation}\label{uvident}
    \bar{u}^{\sigma}_{\bm{p}} \, \gamma^\mu \, v^{\sigma'}_{\bm{p}'} =
    (\bar{v}^{\sigma'}_{\bm{p}'} \, \gamma^\mu \, u^{\sigma}_{\bm{p}})^* ,
\end{equation}
actually are a direct consequence of the self-adjointness of the matrices $\gamma^0 \gamma^\mu$.

\hspace{1cm}


\end{document}